\documentclass[11pt,epsf]{JHEP3}

\def\beq{\begin{eqnarray}}
\def\eeq{\end{eqnarray}}
\def\bea{\begin{eqnarray*}}
\def\eea{\end{eqnarray*}}
\def\beq{\begin{equation}}
\def\eeq{\end{equation}}
\def\ba{\begin{eqnarray}}
\def\ea{\end{eqnarray}}

\def\ov{\overline}

\title{CPT and Other Symmetries in String/M Theory}
\author{Michael Dine \\ 
Santa Cruz Institute for Particle Physics, 
Santa Cruz CA 95064 \\ Email: \email{dine@scipp.ucsc.edu} }
\author{Michael L. Graesser \\ Department of Physics, Caltech, Pasadena, CA 
91125 \\ Email: \email{graesser@theory.caltech.edu}}

\abstract{We initiate a search for non-perturbative consistency 
conditions in M theory.  Some non-perturbative conditions are 
already known in Type I theories; we review these and search for 
others. We focus principally on possible anomalies in discrete 
symmetries.  It is generally believed that discrete symmetries in 
string theories are gauge symmetries, so anomalies would provide 
evidence for inconsistencies. Using the orbifold cosmic string 
construction, we give some evidence that the symmetries we study 
are gauged.  We then search for anomalies in discrete symmetries 
in a variety of models, both with and without supersymmetry. In 
symmetric orbifold models we extend previous searches, and show 
in a variety of examples that all anomalies may be canceled by a 
Green-Schwarz mechanism.  We explore some asymmetric orbifold 
constructions and again find that all anomalies may be canceled 
this way. Then we turn to Type IIB orientifold models where it
is known that even perturbative anomalies are non-universal.  
In the examples we study, by combining geometric
discrete symmetries with continuous gauge symmetries, one
may define non-anomalous discrete symmetries already
in perturbation theory; in other cases,
the anomalies are universal.
Finally, we turn to the question of CPT conservation
in string/M theory.  It is well known that CPT is conserved in all
string perturbation expansions; 
here in a number of examples for which a non-perturbative
formulation is available
we provide evidence that it is conserved
exactly.  }

\keywords{dfs;mxt;shs}
\preprint{CALT-68-2519, SCIPP-04/17}

\begin{document} 

\section{Introduction}
In field theory, the idea that perturbative and non-perturbative
anomalies can render a theory inconsistent is familiar.
In weakly coupled string theory, consistency conditions of various
types are familiar.  For closed strings, modular invariance is
necessary for a unitary, Lorentz-invariant perturbation theory.
Even classical solutions corresponding to smooth manifolds can violate this
condition\cite{wittenargonne}.  For open strings, one has tadpole
conditions.  Non-perturbatively, some examples of consistency
conditions are known in Type I theories\cite{manyauthors}, and
duality considerations suggest\cite{duals} (but do not firmly
establish\cite{aspinwall}) the existence of conditions in some
closed string theories.

Lacking a non-perturbative formulation of string theory, a search
for non-perturbative anomalies involves either study
of topological objects (branes, etc.) or examination 
of features of the low energy theory.  In this note, we adopt
the latter approach, looking at string states which should have a
conventional four dimensional effective field theory description.
We focus, in particular, on the possibility of anomalies in
discrete symmetries.  It is generally believed that discrete
symmetries in string theory are gauged.  This follows from a
general prejudice that global symmetries are implausible in a
theory of gravity.  It is also possible, in many cases, to show
explicitly that these symmetries are remnants of, say, gauge and
general coordinate invariances of a higher dimensional theory.
The most convincing demonstration that any particular symmetry is
a gauge symmetry comes from the study of cosmic string solutions 
\cite{preskill}.
We will discuss aspects of this problem further below.

To convincingly demonstrate an anomaly in such a symmetry
in four dimensions, one
must examine instantons in the low energy theory, and determine if
they violate the symmetry.  One must also check that one cannot
cancel the anomalies by assigning transformation laws to moduli of
the theory, i.e. by a Green-Schwarz mechanism.  The possibility of
such cancellations had been discussed\cite{banksdine} in the
case of smooth Calabi-Yau compactifications, but it actually
follows from an old result of Witten\cite{topologicaltools}.
Witten showed the absence of global anomalies for any configuration
which can be described by the effective ten-dimensional supergravity
theory.  Thus any such anomaly must arise
through some inherently stringy effects.  One might imagine that
already some of the gauge symmetries of orbifold models are
sufficiently ``stringy" that Witten's result might not apply.
However, while extensive
searches among symmetric, supersymmetric
orbifolds\cite{macintire} found many examples of anomalies, all
could be canceled by Green-Schwarz terms.

In the present note we extend this search in two ways.  We
consider non-supersymmetric, symmetric orbifolds.   One might
hope
that, given that non-supersymmetric string theories seem to suffer
from a variety of diseases (particularly tachyons and other
instabilities), that perhaps they would often suffer from discrete
anomalies as well.  However, at least in the small sample of such
orbifolds studied here, this does not occur.

We consider, also, asymmetric orbifolds.  These one might imagine
are more ``stringy" than symmetric orbifolds.  There is not, in
general, an obvious procedure for blowing them up to obtain
theories on smooth manifolds, so they may not fall within the class
of theories considered by Witten.  Here we construct a variety of
asymmetric orbifolds of the heterotic string, both with and without space time
supersymmetry, and do not find anomalies.  

We then search for discrete anomalies in Type IIB orientifold 
constructions. Here we do find non-universal anomalies. In 
contrast to the symmetric and asymmetric orbifolds though, here 
there are gauge singlet massless states which have non-universal 
couplings to gauge fields, already
at tree level. The non-universal discrete anomalies 
may in principle be canceled by assigning a non-linear transformation to 
these fields. The states though, are {\em twisted states}. 

To convincingly demonstrate the {\em absence} of an anomaly, we 
would have to prove the existence of such a non-universal, non-linear 
transformation 
law. 
In the case of anomalies in continuous symmetries, there is 
a one-loop computation one can do to verify the existence of the 
Green-Schwarz term. There isn't something analogous here. Our 
approach is to assume that if the anomaly may be canceled by such a field, 
then it is. The fact that it is always possible to assume that the anomaly 
is canceled this way we take as evidence 
that this view is correct. 

The remainder of the paper investigates a number of M-theory 
backgrounds in which it is possible to
address non-perturbatively the question
of CPT conservation. We study M-theory in 
eleven dimensions, on $S_1/Z_2$, and on $T^3$. At the classical 
level each of these theories has a CPT symmetry. One may ask 
whether this symmetry is preserved by quantum effects. To address 
this, we study matrix model descriptions for these backgrounds
and identify a conserved CPT symmetry. Because the matrix models we study do not 
include all types of five-branes, we cannot argue that this is 
a proof of CPT conservation, but it is certainly strong evidence. 

The paper is organized as follows.  In the next
section, we review some aspects of non-perturbative anomalies in
Type I theories.  We show that orbifold cosmic strings
exist for the cases we study, and explain why this is evidence
(but not proof)
that these symmetries are gauged.
In the section 4, we discuss some symmetric
orbifolds without supersymmetry, and demonstrate the cancellation
of anomalies. Section 5 contains our study of asymmetric
orbifolds.  We
construct the required projectors, and give the spectra for
several models.  The gauge symmetries of the models are often very
intricate, and perturbative anomaly cancellation provides a
non-trivial check on the massless spectrum.  We identify
various discrete symmetries of the models, and compute the
anomalies.  The principal subtlety in this analysis lies in
determining the transformation properties of the twisted ground
states
under the discrete symmetries.  
In section 6 we search for discrete anomalies in Type IIB string 
compactifications with D5 branes located at 
orbifold fixed points. 
The final section 
studies the CPT properties of a few M-theory 
backgrounds and their corresponding matrix 
model descriptions. 

The Appendices provide the details relevant for 
constructing the states 
in both the symmetric and asymmetric orbifolds and the 
Type IIB orientifolds, 
as well as their charges 
under the discrete symmetries of the low-energy theory.

\section{Non-Perturbative Constraints in Type I Theory and Horava-Witten Theory}

It is known that there are non-perturbative consistency conditions
for Type I theories.  Ref. \cite{manyauthors} provides a particularly
simple example of the problem.  When one mods out a weak coupling heterotic
string by some symmetry, one of the conditions is that the action
of the symmetry should be well defined.  In the $O(32)$ heterotic string,
this leads, among other constraints,
to restrictions on the possible twists in order that the
action of the symmetry on fermions and on spinor representations
of $O(32)$ be sensible.  In the Type I theory, however, the
spinor representations of $O(32)$ only arise at the
non-perturbative level.  As a result, some vacua which seem
perfectly sensible in perturbation theory are ill defined
non-perturbatively.  On the weak coupling heterotic side, the
problem is obvious, but on the Type I side, an understanding of
the non-perturbative structure of the theory is required.

This type of argument can be extended to many cases.  For example,
for weak coupling heterotic strings compactified on Calabi-Yau
spaces, there are restrictions on the possible Wilson lines
arising from modular invariance in twisted sectors; the condition
is simply level matching.  On the Type I
side, these constraints can be understood by considering a $D1$
brane wrapped on a non-contractible loop.
Now there is a consistency condition for the existence of this D
brane:  the states of the brane must level match.  This can be
understood by noting that level matching is just the statement
that the state is invariant under shifts of $\sigma$ by $ \pi$,
where now $\sigma$ can be taken as a parameter describing the
wrapped string.  Since the $D1$ string is just the dual of the
heterotic string, this level matching condition is identical to
the level matching condition of the weak coupling
theory.\footnote{We thank Joe Polchinski for a discussion of these
issues.}

We can apply this sort of reasoning to the strong coupling limit
of the $E_8 \times E_8$ theory, i.e. to Horava-Witten theory\cite{horavawitten}.
Again, in the weak coupling limit, one has the constraints of
level matching and consistency of spinor propagation.  In the
strong coupling limit one obtains these requirements by
considering membranes stretched between the two walls and wrapped
around non-contractible loops.  It is interesting that if one does
not impose the level matching condition, one does find discrete
anomalies.

Because we understand in each of these cases the consistency
condition on the weak coupling side, these conditions are not in
some sense new.  A new condition, however, arises in the case of
Horava-Witten theory when one breaks the supersymmetry by imposing
a different chirality condition on each of the two walls.  The
condition is, again, level matching for the wrapped branes; the
condition is new, because we do not know a weakly coupled dual.
\footnote{We thank Tom Banks for this example.}

These examples illustrate that there are non-perturbative
consistency conditions in M theory.  For closed string theories,
however, we do not have examples of non-perturbative consistency
conditions.   Indeed, with the exception of the non-supersymmetric
Horava-Witten example (and perhaps analogous examples
in Type I theories), all of the known examples can be understood
in terms of modular invariance in closed string duals.  It would
be interesting to find examples of inconsistencies, then, in
weakly coupled closed string theories.  Much of the rest of this
note is devoted to such a search.

\section{Discrete Symmetries as Gauge Symmetries:  Cosmic Strings}

It is believed that discrete symmetries in string/M theory are gauged.  Certainly
this can be seen in many examples.
The discrete symmetries of Calabi-Yau spaces are often discrete
subgroups of the full ten-dimensional Lorentz group.  Many duality
symmetries are gauge symmetries, as is the symmetry which exchanges
the two $E_8$'s of the heterotic string.
Perhaps the most conclusive demonstration that a discrete symmetry is
gauged is provided by the presence cosmic strings.  The point is illustrated by
a $U(1)$ theory with a Higgs field of charge $N$.  A Higgs
expectation value breaks $U(1) \rightarrow Z_N$.  Such a theory
admits cosmic strings.  If one brings a field of charge $1$ around
such a string, it picks up a phase $\alpha = e^{2\pi i/N}$.
Similarly, the angular momenta of fields are fractional for such a
field, $m-{1 \over N}$.

One way to test whether discrete symmetries in string theories are
gauge symmetries, then, is to construct cosmic string solutions for
which particles pick  up a suitable phase when they circulate
around the string.   Orbifold cosmic strings\cite{orbifoldstrings}
provide an example of this phenomenon.  Consider a weakly coupled
string orbifold with a $Z_2, Z_3,Z_4$ or $Z_6$ symmetry.  For
example, for the well-known $Z_3$ orbifold, there is, at suitable
points in the moduli space, $Z_6$ symmetries which acts on each
of the orbifold planes.  One can consider compactification of this
theory on an extremely large torus, with a $Z_6$ symmetry, and mod
out by the product of one of the internal symmetries and this
symmetry in what we will call the transverse space.   In the
language of orbifolds, one can think of modding out the torus,
$T_8$, by a product of symmetries, $g$ and $h$, where $g$ is the
usual $Z_3$ and $h$ is the additional $Z_6$.

The fixed points of the symmetry $h$ are naturally thought of as
cosmic strings.  Now consider the various sectors.  First, there
are the sectors {\it untwisted} by $h$.  These sectors are distinguished
by the existence of a set of momenta in the transverse directions.
For a given momentum state, $\vert p \rangle$, one can construct
states transforming as $\alpha^k$, from
\beq
\vert p^{(k)} \rangle = \sum_{r=1}^N \alpha^{rk} h^r \vert p
\rangle ~~~~~
h \vert p^{(k)} \rangle = \alpha^k \vert p^{(k)} \rangle.
\eeq
In general, states in this sector can be written as products,
\beq
\vert \psi \rangle = \vert {\rm internal} \rangle \times \vert p^{(k)}
\rangle.
\eeq
Invariant states are states for which the transformation of the
transverse momentum under $h$ compensates the discrete
transformation of the {\it internal} state.  This is precisely the
effect expected for a cosmic string.

Banks (unpublished) points to other examples.
Consider the $SL(2,Z)$ symmetry of the
Type IIB theory in ten dimensions.  The analog of the cosmic
string, in this case, is a seven brane.  It is easy to see that
$\tau \rightarrow \tau +1$ is connected with the behavior of the
dilaton-axion model as they encircle the seven brane.

\section{A Search for Discrete Gauge Anomalies in Weakly Coupled Heterotic
Strings:  Symmetric Orbifolds With or Without Wilson Lines}
\label{symmorb}

The discussion above suggests that in many -- and possibly all
--cases discrete symmetries in string theory are gauge symmetries,
and any violation of such symmetries would imply an inconsistency.
So it is interesting to search for such violations.  In the past,
limited searches have been conducted among supersymmetric models,
and have failed to find examples of such anomalies\cite{macintire}.  Here we
enlarge the search, including models without supersymmetry.  This
case is particularly interesting, since such models have a number
of troubling features.  They typically contain tachyons, at least
in regions of the moduli space, and they are subject to
catastrophic decay processes\cite{wittenkkdecay,fabingerhorava}.

Before considering weakly coupled closed strings, it is useful to
note that discrete anomalies are closely connected to the
non-perturbative consistency conditions for Type I theory and
Horava-Witten theory which we mentioned earlier.  If we consider
compactification of these theories on a large radius Calabi-Yau,
with Wilson lines, in general we find discrete gauge anomalies,
i.e. we find that instantons in non-Abelian groups in the low
energy theory violate the symmetry.  However, if one imposes the
level-matching conditions, the anomalies cancel.  This point is
illustrated by the case of the quintic in $CP^4$, described at
length in \cite{GSW}.  This theory has a set of $Z_5$
symmetries.  Consider the $O(32)$ theory, for example, and
includes a Wilson line $(1,1,0...\dots 0)$, where the notation
means (in the fermionic formulation) group the fermions into
sixteen complex pairs, and twist the first two by a $Z_5$ phase,
while  leaving the others alone.  This particular choice of Wilson
lines does not level match.  In this case one finds that the
discrete anomalies in the low energy groups are not the same.
Similar examples arise in the $E_8
\times E_8$ theory.

\subsection{Discrete Symmetries}

In this section, we focus on discrete
symmetries of orbifold compactifications
to four dimensions.  We will limit
our considerations to cases where
the original torus is a product of three
two-dimensional tori. In general, these models have a number 
of discrete symmetries. The orbifold group removes one linear 
combination of symmetries of the lattice, but others survive and 
are potentially anomalous. These are the quantum $Z_N$ symmetry 
\cite{quantumvafa}, and for toroidal compactifications, point 
group symmetries that are independent rotations of two of the 
three tori. 

The quantum symmetry 
is the automorphism group of the orbifold group that for
a $Z_N$ orbifold is
the $Z_N$ cyclic permutation of its elements. Untwisted
states are neutral, whereas states in the
$k$-th twist sector have charge $k/N$. String
interactions conserve this symmetry.

\subsection{Models}

Perhaps the simplest models to study in a search for such
anomalies are symmetric orbifolds of the heterotic string.
$Z_6$ orbifolds provide examples which are chiral
and break supersymmetry.  The $Z_6$ symmetry
can be realized by a simple product of two dimensional lattices.
At such points in moduli space, the models have 3 $Z_6$ symmetries
(and possibly additional symmetries, such as permutations).
One can mod out by one
combination of symmetries; others survive, and are potentially
anomalous.  We have considered several variations on the $Z_6$
orbifold, with and without supersymmetry.  In each case, we find
that the anomalies cancel.

\subsubsection{Supersymmetric $Z_6$}

A supersymmetric $Z_6$ twist, satisfying level matching is
\beq
\phi =  \frac{1}{6}(1,1,-2)
\eeq
Choosing the standard embedding for the gauge
shift breaks $SO(32) \rightarrow SU(2) \times U(1) \times SO(26) $.
The massless fermions are given 
in Table \ref{susyz6spec}.
The table
also lists the discrete charges of the
states under the two $Z_6$ symmetries that remain after
orbifolding. Here $\gamma$ corresponds to rotations of the
third plane, whereas
$\eta$
corresponds to simultaneous $Z_6$ rotations of the first two
planes.
All the symmetries are anomaly-free. One also finds 
that the quantum $Z_6$ symmetry is not anomalous (not shown).

\small
\TABLE[h]{
\begin{tabular}{|c|c|c|c|} \hline
sector & $~G~ =~ SO(26) \times SU(2) \times U(1)^2
$ & $Z_6$ & $Z^{\prime}_6$  \\ \hline
& & & \\
untwisted & $({\bf 325},{\bf 1};0,0;\eta,\gamma^{1/2})+ 
({\bf 1},{\bf 3};0,0,\eta,\gamma^{1/2}) $ & & \\
  & $({\bf 26},{\bf 1};-1,-1; \eta^{-1}, \gamma^{\frac{1}{2}})+
 2({\bf 1},{\bf 2};2,1;1,\gamma^{-\frac{1}{2}})+
2  ({\bf 26},{\bf 2}; -1,0; 1,
\gamma^{-\frac{1}{2}})~$ & $(\gamma^3,\gamma^5) $ & $(\eta^4,\eta^4)$ \\
& &  & \\ \hline
& & & \\
 $n=1$  & $6 ({\bf 1},{\bf 2}; 0,-\frac{2}{3};
\eta^{-\frac{5}{3}}, \gamma^{\frac{1}{6}})+
3({\bf 26},{\bf 1};-1,-\frac{2}{3}; \eta^{-\frac{2}{3}},
\gamma^{\frac{1}{6}})   $ &  $(\gamma,\gamma) $ & $(\eta^2,\eta^2)$ \\
& & & \\ \hline
& & & \\
$n=2$        & $10({\bf 26},{\bf
1};-1,-\frac{1}{3};\eta^{-\frac{1}{3}},
\gamma^{-\frac{1}{6}})+5 ({\bf 26},{\bf 1};-1,-\frac{1}{3};
\eta^{-\frac{1}{3}}, \gamma^{\frac{17}{6}}) $ & & \\
& $+20({\bf 1},{\bf 2};0,-\frac{1}{3}; \eta^{-\frac{4}{3}},
\gamma^{-\frac{1}{6}})+10({\bf 1},{\bf 2} ;0,-\frac{1}{3};
\eta^{-\frac{4}{3}}, \gamma^{\frac{17}{6}})$  &  $(\gamma,\gamma^5) 
$ & $(\eta^2,\eta^4)$  \\
& $+8({\bf 1},{\bf 2};0,-\frac{1}{3};
 \eta^{\frac{8}{3}}, \gamma^{-\frac{7}{6}})+
4({\bf 1},{\bf 2};0,-\frac{1}{3}; \eta^{\frac{8}{3}},
\gamma^{\frac{11}{6}})$ & & \\
& & & \\ \hline
& & & \\
$n=3$        & $6({\bf 26},{\bf 1};-1,0; 1,
\gamma^{-\frac{1}{2}})+5({\bf 26},{\bf 1};1,0;\eta^{4},
\gamma^{-\frac{1}{2}}) $  &  $(\gamma,\gamma) $ & $(\eta^4,\eta^2)$ \\
& $+12({\bf 1},{\bf 2};0,0;\eta^{-1},
\gamma^{-\frac{1}{2}} )+10 ({\bf 1},{\bf 2};0,0; \eta^5,
 \gamma^{-\frac{1}{2}})$  & & \\
& & & \\ \hline
& & & \\
total & &$(1,1)$  & $(1,1)$ \\ 
& & & \\ \hline 
\end{tabular}
\label{susyz6spec}
\caption{Supersymmetric symmetric $Z_6$ orbifold with standard embedding.
Here $\gamma$ and
$\eta$ correspond to $Z_6$ rotations of the third torus and
simultaneous $Z^{\prime}_6$ rotations of the first and second
torus, with $\gamma^6=\eta^6=1$. $Z_6$ and $Z^{\prime}_6$ 
discrete gauge anomalies are indicated.}}
\normalsize

\subsubsection{Nonsupersymmetric $Z_6$ model with Wilson lines}

A {\it non-supersymmetric} symmetric
$Z_6$ orbifold with Wilson lines, satisfying the
above level matching conditions is given by
\beq
\phi  = \frac{1}{6}(0,1,1,4) ~,~
\beta = {1 \over 6}(1,1,4;0^{13}) ~, ~
a_{5} =-a_{6} = \frac{1}{6}(0^3,4^6,0^7) ~.
\eeq
The standard embedding for $\beta $ combined with the
choice of Wilson line breaks
$SO(32) \rightarrow SO(14)
\times SU(6) \times SU(2) \times U(1) ^3$.
The states for this model are given in Table \ref{nonsusyz6spec}.
In the $n$-th twisted sector there will also
be a Wilson line for each fixed point, and in general
the states at each fixed point will be different.
The third plane, which has the Wilson line,
has 3 fixed points for $n=1$
and $2$, and no fixed points for $n=3$.
In Table \ref{nonsusyz6spec} the integer $m=0$,$1$ and $2$
refer to these
fixed points. It
also lists the discrete charges of the
states under the two $Z_6$ symmetries that remain after
orbifolding. Here $\gamma$ corresponds to rotations of the
third plane, whereas
$\eta$
corresponds to simultaneous $Z_6$ rotations of the first two
planes.

\footnotesize 
\TABLE[h]{
\begin{tabular}{|c|c|c|c|} \hline
$(n,m)$ & $~G = SO(14) \times SU(6) \times
SU(2) \times U(1)^3 $ & $Z_6$ & $Z^{\prime}_6$ \\ \hline
&  & & \\
$(0,0)$ & $2 ({\bf 14},{\bf 1},{\bf 1};0,-1,0;\gamma^{-\frac{1}{2}},
1)+
({\bf 14}, {\bf 1},{\bf 2};-1,0,0; \gamma^{\frac{1}{2}}, \eta^{5}) $
& $(1,1,\gamma^{5/2})$ & $(\eta^{2},1,\eta^{5})$ \\
 & $+({\bf 1},{\bf 1},{\bf 2}; \mp 1, \pm 1,0;
\gamma^{\frac{1}{2}}, \eta)+({\bf 1},{\bf 1},{\bf 2};
1,1,0;\gamma^{\frac{1}{2}},\eta^{5})$ & & \\
& & & \\ \hline
& & & \\
$(2,0)$ & $4  ({\bf 14},{\bf 1},{\bf 1};-\frac{2}{3},-\frac{1}{3},0;
 \gamma^{-\frac{1}{6}}, \eta^{8/3})
+4  ({\bf 1},{\bf 6},{\bf 1};-\frac{2}{3},-\frac{1}{3},1;
\gamma^{-\frac{1}{6}}, \eta^{8/3})$ & & \\
& $+4  ({\bf 1}, {\bf \ov{6}},{\bf 1};-\frac{2}{3},
-\frac{1}{3},-1; \gamma^{-\frac{1}{6}}, \eta^{8/3})+
8  ({\bf 1},{\bf 1},{\bf 2};\frac{1}{3},-\frac{1}{3},0;
\gamma^{-\frac{1}{6}}, \eta^{5/3})$ & $(\gamma^{14/3},
\gamma^{14/3},\gamma^{29/6})$ & $(\eta^{10/3},\eta^{10/3},\eta^{17/3})$ \\
& $+5  ({\bf 1},{\bf 1},{\bf 2};\frac{1}{3},-\frac{1}{3},0;
\gamma^{-\frac{7}{6}}, \eta^{-1/3})$  & & \\
$(2,1)$ & $5 ({\bf 1},{\bf 1},{\bf 2};\frac{1}{3},-\frac{1}{3},2;
\gamma^{-\frac{1}{6}}, \eta^{-\frac{1}{3}})+
4  ({\bf 1},{\bf \ov{6}},{\bf 1};-\frac{2}{3},-\frac{1}{3},1;
\gamma^{-\frac{1}{6}}, \eta^{8/3})$ & 
$(1,\gamma^{-2/3},\gamma^{-5/6})$ & $(1,\eta^{-4/3},\eta^{-5/3})$ \\
$(2,2)$ & $5  ({\bf 1},{\bf 1},{\bf 2}; \frac{1}{3}, -\frac{1}{3}, -2;
\gamma^{-\frac{1}{6}}, \eta^{-\frac{1}{3}})+
4 ({\bf 1},{\bf 6},{\bf 1}; -\frac{2}{3}, -\frac{1}{3}, -1;
\gamma^{-\frac{1}{6}}, \eta^{8/3})$ &$(1,\gamma^{-2/3},
\gamma^{-5/6})$  & $(1,\eta^{-4/3},\eta^{-5/3})$\\
& & & \\ \hline
& & & \\
$n=3$ & $5  ({\bf 1},{\bf 1},{\bf 2};0,1,0;
\gamma^{-\frac{1}{2}}, \eta^2)+6  ({\bf 1},{\bf 1},{\bf 2};
0,-1,0; \gamma^{-\frac{1}{2}},1)$ & $(\gamma^4,1,\gamma^{9/2})$ 
& $(\eta^2,1,\eta^2)$ \\
& $+5  ({\bf 14},{\bf 1},{\bf 2}; 0,0,0;
\gamma^{-\frac{1}{2}}, \eta^4)$ & &  \\
& & & \\ \hline
& & & \\
$(5,0)$ & $2  ({\bf 14},{\bf 1},{\bf
1};\frac{1}{3},-\frac{1}{3},0;\gamma^{-\frac{1}{6}},
\beta^{\frac{5}{3}})+2\times  ({\bf 1},{\bf \ov{6}},{\bf
1};\frac{1}{3},-\frac{1}{3},-1;\gamma^{-\frac{1}{6}},
\eta^{5/3})$ & & \\
 & $2  ({\bf 1},{\bf {6}},{\bf
1};\frac{1}{3},-\frac{1}{3},1;
\gamma^{-\frac{1}{6}}, \eta^{\frac{5}{3}})+3\times ({\bf 1},{\bf
1},{\bf 2};-\frac{2}{3},-\frac{1}{3};0,\gamma^{-\frac{1}{6}},
\eta^{8/3}) $ & $(\gamma^{16/3},\gamma^{16/3},\gamma^{25/6})$ & 
$(\eta^{2/3},\eta^{2/3},\eta^{-8/3})$ \\
& $+({\bf 1},{\bf 1},{\bf
2};\frac{4}{3},-\frac{1}{3},0;\gamma^{-\frac{1}{6}},
\eta^{\frac{2}{3}})+({\bf 1},{\bf 1},{\bf
2};-\frac{2}{3},-\frac{1}{3},0;\gamma^{-\frac{7}{6}},
\eta^{\frac{2}{3}})$ & & \\
& & & \\
& &  & \\
$(5,1)$ & $2  ({\bf 1},{\bf \ov{6}},{\bf
1};\frac{1}{3},-\frac{1}{3},1;\gamma^{-\frac{1}{6}},
\eta^{\frac{5}{3}}) +({\bf 1},{\bf 1},{\bf
2};-\frac{2}{3},-\frac{1}{3},2;\gamma^{-\frac{1}{6}},\eta^{\frac{2}{3}})$
& $(1,\gamma^{-1/3},\gamma^{-1/6})$ & $(1,\eta^{10/3},\eta^{2/3})$ \\
$(5,2)$ & $2 ({\bf 1},{\bf 6},{\bf
1};\frac{1}{3},-\frac{1}{3},-1;\gamma^{-\frac{1}{6}},
\eta^{\frac{5}{3}})+({\bf 1},{\bf 1},{\bf
2};-\frac{2}{3},-\frac{1}{3},-2;\gamma^{-\frac{1}{6}},
\eta^{\frac{2}{3}})$ & $(1,\gamma^{-1/3},\gamma^{-1/6})$ & 
$(1,\eta^{10/3},\eta^{2/3})$ \\
& & & \\ \hline
& & & \\
total & & $(\gamma^2,\gamma^2,\gamma^2)$ & $(\eta^2,\eta^2,\eta^2) $ 
\\ 
& & & \\ \hline  
\end{tabular}
\label{nonsusyz6spec}
\caption{Nonsupersymmetric symmetric
$Z_6$ model with the standard embedding 
and Wilson lines. Here $\gamma$ and
$\eta$ correspond to $Z_6$ rotations of the third torus and
simultaneous $Z^{\prime}_6$ rotations of the first and second
torus, with $\gamma^6=\eta^6=1$. $Z_6$ and $Z^{\prime}_6$ 
discrete gauge anomalies are indicated. }}
\normalsize

The discrete anomalies from each sector are found in the last two 
columns of each  table. One finds that the discrete anomalies 
are universal, and so may be canceled by a discrete Green-Schwarz 
mechanism. Note that the untwisted and twisted sectors each 
have non-universal anomalies and it is only their total that is 
universal. This feature is typical of the models we study. The 
quantum twist symmetry is also non-anomalous, although it too 
receives contributions from all twisted sectors (not shown).

\section{A Search for Discrete Gauge Anomalies in Weakly Coupled Heterotic
Strings:  Asymmetric Orbifolds}

Given Witten's results, one might have suspected that one would
not find anomalies in symmetric orbifold models.  In the
supersymmetric case, these models can be blown up to smooth
Calabi-Yau manifolds, and no anomalies can appear in this limit.
In the non-supersymmetric cases, the situation is less clear, since one
does not have an argument that one can find solutions of the classical
equations corresponding to smooth manifolds, but
one might still suspect that these models are not so different
from the smooth cases, and that this is why we fail to find
anomalies.  Asymmetric orbifolds\cite{ao}, one might hope, are more likely
``stringy," and results obtained by considering smooth geometries
might not hold.
In this section, we search for (but do not find) discrete anomalies in
asymmetric orbifold models, both with and without supersymmetry
(in the weak coupling limit).

The asymmetric orbifold construction has been developed by a
number of authors.  We follow, particularly, the work of
\cite{ibanez} and \cite{kt}.  The procedure is
straightforward.  We start with a toroidal compactification of the
heterotic string, on a lattice with a set of unbroken $T$-duality
symmetries.  We mod out this theory by a subgroup of this
symmetry, and add twisted sectors so as to obtain a
modular-invariant partition function.  The result of this
construction can be expressed in terms of a set of projectors for
each twisted sector.  We then construct the massless spectrum.
Typically, perturbative anomaly cancellation provides a highly
non-trivial check on the construction.  Then we examine the
discrete symmetries of the orbifold theory, and determine the
transformation properties of the states under these symmetries.
This is slightly non-trivial, since the twisted ground states
themselves transform.  With this information in hand, it is
straightforward to ask whether instantons of the low energy theory
violate the discrete symmetry.

\subsection{Nonsupersymmetric Examples}

We first consider a set of non-supersymmetric orbifolds.  As in
the case of symmetric orbifolds above, we will not worry whether
the states have tachyons (in the symmetric cases, anomaly
cancellations occurred even with tachyons).  

The fields in the theory are 
the 16 freely
interacting left--moving (LM) real scalars $H^a_L$, 
three LM complex scalars $X^i_L$, three right--moving 
(RM) complex scalars 
$X^i_R$ and their fermionic partners 
$\tilde{\psi}_i$. The scalars $X$ cannot be interpreted 
as describing the coordinates of an internal 
manifold as the left and right movers are 
treated differently. 

For all of these
orbifolds, we will take the underlying lattice to be
\beq
\Gamma_{(16)} \times \Gamma_{(4,4)}(D_4)
\times \Gamma_{(2,2)}(A_2) ~.
\label{latticex}
\eeq

The lattice $\Gamma_{(2,2)}(A_2)$ is formed from the 
$SU(3)$ lie algebra in the manner described in Appendix B.
The weight lattice is
generated by an
element $e_1$ from the ${\bf 3}$, and an element $e_2$
from the
$\overline{\bf
3}$. It is important to note that
the weight lattice has both an obvious $Z_3$ and a less obvious
$Z_6$
automorphism symmetry.
The $Z_3$ is a Weyl symmetry of the weight lattice, corresponding
to a discrete $SU(3)$ rotation that permutes the elements of the
${\bf 3}$. The Lorentzian lattice $\Gamma_{(2,2)}(A_2)$ has a
separate left and right $Z_3$ symmetry, since this automorphism
preserves the condition $p_L - p_R \in \Lambda_R$. The $Z_6$ symmetry
is also an automorphism of the $SU(3)$ lattice, but it corresponds
to a $Z_6$ rotation of $e_1 \rightarrow e_2$, $e_2 \rightarrow
-e_1$, which is a mapping between the
conjugacy classes. Consequently, this is a symmetry
of $\Gamma_{(2,2)}(A_2)$ only if it acts simultaneously on the left and
the right.

The $\Gamma_{(4,4)}(D_4)$ lattice is formed from the $SO(8)$
lie algebra as described in Appendix B. 
The weight lattice has four conjugacy
classes,
the root, the vector and the two spinorial classes.
These can be generated by four elements; one element
may be chosen from
the
spinorial class, and the other three can be taken from the vector class.
The $SO(8)$ lattice has a $Z_6$ Weyl symmetry \cite{bouwknegt}.
Here we focus on the obvious
$Z_2$ Weyl symmetry that sends
a weight $w \rightarrow -w$. The lattice also has a $S_3$ permutation
symmetry, which is just triality.

To construct an asymmetric orbifold we divide by
a subset of these Weyl symmetries; in particular
the $Z_3$ and $Z_2$.
For
the right movers, in each case, we will take the projector to be
the $Z_6$ twist
\beq
X^i _R  \rightarrow X^i _R  e^{2 \pi i \phi^i _R}
\eeq
where
\beq
 \phi _R = \left({1 \over 3}, {1 \over 2},{1 \over 2} \right), 
\eeq
describing a $Z_3$ and $Z_2$ twist 
in the two and four dimensional lattices, respectively. 
This twist leaves no unbroken supersymmetries. 
The RM weights $r$, describing the
positive helicity
massless fermions for the different sectors
are given in the Table \ref{nonsusyfermi}.(Both states with a 
`$\pm$' satisfy the projector and are included in the spectrum.) 
For the left movers
we choose $\phi_L=0$.

One may check
that this twist satisfies the second and last
level matching conditions (\ref{lm1})
and (\ref{lm2}). The details for satisfying 
(\ref{lm2}) are provided in 
Appendix B.

On the left, we will take the group action to be a shift by a
vector $\beta_L$ and, as already stated, $\phi_L=0$.
Our different models will be
characterized
by different choices of $\beta_L$ that satisfy the level matching
conditions (\ref{lm1}) and (\ref{lm2}).

\TABLE[h]{\begin{tabular}{|c|c|} \hline
 sector  & ~{\rm Ramond state}  \\ \hline
& \\
   & $ ~r_1= (\frac{1}{2}, \frac{1}{2},
\pm \frac{1}{2}, \pm \frac{1}{2}) $
 \\
untwisted & \\
& $ r_2 = ~(\frac{1}{2},- \frac{1}{2},
\pm \frac{1}{2}, \mp \frac{1}{2})$ \\
& \\ \hline
$n=1$  & ~{\rm none }  \\ \hline
& \\
$n=2$ & $~( \frac{1}{2}, - \frac{1}{2},
- \frac{3}{2}, -\frac{1}{2} )
~+~ ( \frac{1}{2}, - \frac{1}{2}, - \frac{1}{2},
-\frac{3}{2})$ \\
& \\ \hline
& \\
 $ n=3 $ & $~( \frac{1}{2}, - \frac{3}{2},
- \frac{3}{2}, -\frac{3}{2} )$  \\
& \\ \hline
& \\
$n=4$ & $~( \frac{1}{2}, - \frac{3}{2},
- \frac{3}{2}, -\frac{3}{2} ) ~+~
( \frac{1}{2}, - \frac{3}{2},
- \frac{5}{2}, -\frac{5}{2} )$  \\
& \\ \hline
& \\
$n=5$ & $~(\frac{1}{2}, -\frac{3}{2},
- \frac{5}{2}, -\frac{5}{2})$  \\
& \\ \hline
\end{tabular}
\caption{Non-supersymmetric $Z_6$: Positive--helicity RM Ramond sector
weights $(r)$. The ground states are given by
$r + n \phi_R$.}\label{nonsusyfermi}}

\subsubsection{Projectors}

It is straightforward to work out
the projectors in the different sectors.  In doing so, the
character transformation formulae (see, for example,
\cite{kt}), are useful.
Performing so-called $S$ and $T$ transformations repeatedly, one obtains
the partition functions.
To obtain the projectors for the massless fermions, we note
that the RM Ramond ground state is already massless, which
has $p_R =0$. In addition, for these models 
all the 
charged states do not contain any oscillators, 
so the oscillator contribution to the 
projectors are not included in the expressions 
below, although it is 
easy to add it back in. 
The projectors
in the nth-twisted sectors
for the massless states are then readily obtained
(these results are special cases of the
formulas in \cite{kt} and elsewhere) : 
\ba
n=0: & ~ & p_L \cdot \beta_L - r \cdot \phi_R = 0 ~\hbox{mod 1},
\nonumber \\
n=1: & ~ & (p_L+\beta_L) \cdot \beta_L - \frac{1}{2} \beta^2 _L
-r \cdot \phi_R -\sum_i \frac{\phi^i_R}{2}
+\left \{ \frac{1}{3} ~\hbox{if}  ~ p_{(2),L} \in  Y_{i=1,2} ~
~\hbox{or} ~0 ~\hbox{if} ~ p_{(2),L} \in  Y_0 \right \} \nonumber \\
& & +\left \{ \frac{1}{2}  ~\hbox{if}
~ p_{(4),L}\in Z_{i=1,2,3} ~\hbox{or} ~0 ~ \hbox{if}
~ p_{(4),L} \in  Z_0 \right \}
= 0 ~\hbox{mod 1} ~, \nonumber \\
n=2: & ~ & (p_L+2 \beta_L) \cdot \beta_L - \beta^2 _L
-r \cdot \phi_R -\sum_i \frac{\phi^i _R}{2} \nonumber \\
& & + \left \{ \frac{2}{3} ~ \hbox{if}  ~ p_{(2),L}
\in  Y_{i=1,2} ~\hbox{or}~
0 ~\hbox{if}~  p_{(2),L}  \in  Y_0 \right \}
= 0 ~\hbox{mod 1} ~, ~p_{(4),L} \in Z_0 ~, \nonumber \\
n=3: & ~ & (p_L+3 \beta_L) \cdot \beta_L - \frac{3}{2} \beta^2 _L
-r \cdot \phi_R- \sum_i \frac{\phi^i _R}{2}+ \frac{1}{2} -
\frac{1}{2}\phi_R^1
\nonumber \\
& & + \left \{ \frac{1}{2} ~ \hbox{if} ~ p_{(4),L} \in  Z_{i=1,2,3} ~
\hbox{or} ~
0 ~ \hbox{if} ~ p_{(4),L} \in  Z_0 \right \}
 = 0 ~\hbox{mod 1} ~, ~ p_{(2),L} \in Y_0 ~.
\label{asymprojs}
\ea
Here $Y_0$ denotes the
$SU(3)$ root lattice, and $Y_{1,2}$ denote the
lattices
generated by the
two fundamental weights.
Similarly, $Z_0$ is the root
lattice of $SO(8)$, and $Z_{1,2,3}$ are the vector and
two spinorial lattices respectively.
The bracket notation means that there is an additional
phase that depends on the weight momentum.
For example, in the $n=1$ twisted
sector
the value $1/2$ is added to the projector
if $p_{(4),L} \in Z_{i=1,2,3}$, whereas no factor (zero)
is added if $p_{(4),L} \in Z_0$. In a symmetric orbifolds
these phases do not appear
as they cancel between the left and right movers.
Finally, the states in the $n=4$ and $n=5$ sectors
are obtained from the CPT conjugates of the states in the
$n=2$ and $n=1$ sectors.

To complete the spectrum the bosonic and fermionic 
degeneracies for each sector must be specified, 
with the latter trivially
obtained from Table \ref{nonsusyfermi}.

The
bosonic
degeneracy may be obtained either directly from the partition
function or from the general formula presented in \cite{ao}.
The partition function 
implies that
in the $n=1$, $n=3$ and $n=5$ sectors the bosonic degeneracy
is $D=2$,
whereas in the $n=2$ and $n=4$ sectors it is
$D=1$. The general formula \cite{ao}
for the bosonic degeneracy of the $n$--th twisted sector is
\beq
D_n= { \prod_i  2 \sin \pi n \phi^i_R \over \hbox{vol}(I_n)} ~,
\label{bosedeg}
\eeq
where vol$(I_n)$ is the volume of the fundamental region of
the lattice left invariant
by the twist, and the product is over only non-vanishing
twists. This gives $D_1=2$, $D_2=1$ and $D_3=2$, which agrees
with the results obtained directly from the partition 
function.

\subsection{Discrete Symmetries and Selection Rules}

Here we discuss some of the discrete symmetries that
exist in the orbifolded theories. To compute 
the discrete anomalies, it is crucial to obtain 
the correct charges. These 
are rather straightforward to obtain for untwisted 
states and also for worldsheet fermions, untwisted or 
twisted, since there exists an explicit 
construction for twisted fermion vertex operators. 
This is discussed in subsection \ref{fermicharge}. 

The main subtlety is with the twisted bosons, 
and in particular the bosonic twist operator. 
In a symmetric orbifold 
this charge is readily obtained from the geometry. But for an 
asymmetric orbifold such geometric intuition is lacking, 
and we must resort to algebraic methods. 
Sections \ref{bosonictwist1} and \ref{bosonictwist2} 
compute this charge using two independent 
methods, which are found to agree.  

To begin though, the asymmetric orbifolds 
considered here have the following abelian
discrete symmetries :

$\bullet$ Quantum $Z_N$ symmetry,
where $N$ is the order of the orbifold group \cite{quantumvafa}.
Only twisted states are charged under this symmetry.
Their
charge is simply $n/N$, where $n$ refers to the $n$--th twisted
sector.

In a symmetric orbifold the existence of the quantum symmetry is
not hard to see. This is because
a twisted string joined to
a number of twisted strings may only form
an untwisted string if the net twist is a multiple of $N$.
For asymmetric orbifolds this argument is too naive, since
here there is no geometric picture
for the fixed points.

To see the existence of a quantum $Z_N$ symmetry 
in asymmetric orbifolds it is simplest to consider 
a correlation function involving a number of fermionic twist
operators,  
$\tau_R$ 
(including excited states) and untwisted fermions,
\beq
\langle e^{-i r_1 \cdot H_R(z_1)} \cdots
e^{-i r_l \cdot H_R(z_k)} \tau^{(l_{1})} _R
(z_{l_1})
\cdots  \tau^{(l_p)} _R (z_{l_p}) \rangle  ~.
\eeq
By explicitly evaluating this correlation 
(see for instance, \cite{polchinskibook}), one finds it 
is non--vanishing only if the sum of the momenta for all
the operators, twisted and untwisted,
vanishes. Now the $i$
fermionic twist operator 
in the $n_i-$th twisted sector has momentum
$r^a+ n_{i} \phi^a_R$
where $\phi^a_R =k^a/N$ are the twists in the 
$n=1$ sector, and at least one $k^a$ and $N$ are 
relatively coprime. Then the total 
momenta from the
twisted sectors can only be canceled by untwisted
states if
$\sum_i n_{i} k^a/N$ is
an integer. This is just the $Z_N$ selection rule described above.

$\bullet$ Discrete
$Z_k$ symmetries of the lattice $\Gamma_{(d,d)}$ that
remain after the orbifold projection. These are
symmetries of the action and stress--tensor
(or the Virasoro algebra) and are therefore
symmetries of the
perturbative string theory.
These may act symmetrically or asymmetrically on $p_L$ and $p_R$.
For instance, the $\Gamma_{(2,2)}(A_2)$ lattice has the
asymmetric
$Z_3$ symmetry $(p_L,p_R) \rightarrow (p_L,\alpha p_R)$, $\alpha^3=1$,
and the symmetric $Z_6$ symmetry $(p_L,p_R) \rightarrow (\gamma p_L,
\gamma p_R)$
where $\gamma^6=1$. The $\Gamma_{(4,4)}(D_2)$ lattice 
has a $Z_2$ symmetry that acts asymmetrically. 

All these symmetries provide selection
rules for correlation functions. One can ask whether
these symmetries are broken by instantons in the
low--energy theory. More concretely, we compute 
the variation of the 't Hooft operator for the 
each of the gauge groups, and look to see if the 
variations are all universal. 
To do this, we need the discrete charges of 
the spacetime zero modes, and  
the number of zero modes
for each representation in the one instanton background. 
The latter 
may be found in Table \ref{zeromodes}.
To compute the charges of the spacetime zero modes 
we need the charges of 
the worldsheet states. For worldsheet fermions this 
is described in subsection \ref{fermicharge} and for worldsheet 
bosons in subsections \ref{bosonictwist1} and 
\ref{bosonictwist2}. Again, the main subtlety is finding the 
charge of the twisted bosonic ground state. 

We find that these symmetries are often anomalous,
but since the anomalies are always universal they
may be canceled by a discrete Green--Schwarz mechanism.
In subsections \ref{othermoduli} and \ref{othersusymoduli} 
we explain why a non-universal 
anomaly cannot be canceled, despite the existence of 
other massless scalars. 

\TABLE[h]{
\begin{tabular}{|c|c|c|}  \hline
$G$ & $R$ & $n$ \\ \hline
 & fund & $ 1 $  \\
$SU(N)$ & anti-sym  & $N-2 $  \\
  & adjoint & $2N$ \\ \hline
  &  vector  & $2$  \\ \
$SO(2N>4)$ & spinor & $2^{N-3}$    \\
 & adjoint & $ 4N -4 $ \\ \hline
$SO(4)$ & spinor & $1$ \\ \hline
\end{tabular}
\caption{Instanton zero mode counting.}
\label{zeromodes}}

\subsection{Models}

Here we give in some detail the massless spectra of some models,
and show that the anomalies are universal.
These models are specified by a choice of a left-moving
shift $\beta_L$ that satisfies level matching. The massless
fermions are then obtained from the projectors provided in the
previous section. Their charge under the discrete $Z_3$ is
obtained from the rules described in the
Appendix C.

In all of these models
the perturbative, $SU(2)$ Witten,
quantum $Z_N$ and discrete $Z_3$ anomalies either vanish, or may be canceled
by a Green-Schwarz mechanism. It is interesting to
note that the $Z_3$ anomalies are not universal
in the untwisted sector, and are only
universal after including
the twisted sectors.

\subsubsection{Other moduli?} 
\label{othermoduli}

We first ask:  are there other moduli besides the dilaton
which could have canceled non-universal anomalies?
But in these examples, it is
easy to see that the answer is no.  The point is that there are simply
no other massless scalars, apart from the dilaton, which are neutral under
all symmetries.  First, in the untwisted
sector, the asymmetric
twist projects out all of the geometric moduli.  In
the twisted sectors, because of the shifts all
massless states are charged under the gauge symmetries.  They 
are also charged under the quantum $Z_N$ symmetry. 
Thus, even
if there are massless bosons in the twisted sectors, and some of
these are moduli, they cannot couple to $F \tilde F$, and thus
could not have played a role in anomaly cancellation.

\TABLE[h]{
\begin{tabular}{|c|c|c|} \hline
sector & $~SU(4)\times SO(22)\times SU(3)'\times
SO(6) \times U(1)^3 $ & $Z_3$ anomaly \\ \hline
& & \\
& $~ 2 \times ({\bf 4} ,{\bf 22} ,{\bf 1} ,{\bf 1} ; -1,0,0;r_2) +
2 \times ({\bf 6} ,{\bf 1},{\bf 1} ,{\bf 1} ; -2,0,0;r_1)$ &  \\
 untwisted &
$~2 \times  ({\bf \overline{4}},{\bf 1},{\bf 1},{\bf 1}; 1,-1,0;r_2) +
2 \times ({\bf 1},{\bf 22},{\bf 1},{\bf 1} ; 0,-1,0;r_1) $ &
$(1,1,1,\alpha)$ \\
   & $~ 2 \times ({\bf 1},{\bf 1},{\bf 1},{\bf 6} ; 0,0,-1;r_1) $ & \\
& & \\ \hline
 $n=1$  & ~ no states  & \\ \hline
& & \\
$n=2$ & $~ 2 \times
({\bf 4} ,{\bf 1} ,{\bf 3}  \hbox{ and }
{\bf \overline{3}},{\bf 1} ; \frac{1}{3},-\frac{1}{3},\frac{2}{3})
+ 2 \times
({\bf 4} ,{\bf 1} ,{\bf 1} ,{\bf 6};
\frac{1}{3},-\frac{1}{3},-\frac{1}{3})$ & \\
 & $~2 \times
({\bf \overline{4}},{\bf 1} ,{\bf 1} ,{\bf 1}
;-\frac{5}{3},-\frac{1}{3},\frac{2}{3})$ & $(1, 1, \alpha,\alpha^{2})$ \\
& & \\ \hline 
& & \\ 
& $~2 \times
({\bf 6} ,{\bf 1} ,{\bf 1} ,{\bf 1} ;0,1,0) + 2 \times
({\bf \overline{4}},{\bf 1},{\bf 1},{\bf 6}; -1,0,0)$ & \\
$n=3$ & $~2 \times ({\bf 4},{\bf 1},{\bf 1},{\bf 1};1,0,-1) + 2 \times
({\bf 4},{\bf 1},{\bf 1},{\bf 4}_S \hbox{ and } {\bf 4}_{S'};
1,0,\frac{1}{2})$ & $(1,\alpha^2,1,\alpha^2)$ \\
 & $~2 \times ({\bf 1},{\bf 22},{\bf 1},{\bf 1}; 2,0,0) + 2 \times
({\bf 1},{\bf 1},{\bf 1},{\bf 1}; -2,-1,0)$  &\\
& & \\ \hline
& & \\
$n=4$ & $~2 \times
({\bf 1},{\bf 1},{\bf 3} \hbox{ and } {\bf \overline{3}},{\bf 1};
-\frac{4}{3},- \frac{2}{3},-\frac{2}{3}) +
 2 \times
({\bf 1},{\bf 1},{\bf 1},{\bf 6};
-\frac{4}{3},- \frac{2}{3},\frac{1}{3})$ &
$(\alpha,\alpha,\alpha^2,\alpha)$ \\
 & $~2 \times
({\bf 1},{\bf 22},{\bf 1} ,{\bf 1};
-\frac{4}{3}, \frac{1}{3},-\frac{2}{3}) + 2 \times
({\bf \overline{6}},{\bf 1},{\bf 1},{\bf 1};
\frac{2}{3}, -\frac{2}{3},-\frac{2}{3})$ & \\
& & \\ \hline
& & \\
$n=5 $ &  $~2 \times
({\bf 1},{\bf 22},{\bf 1},{\bf 1};
-\frac{2}{3}, \frac{2}{3},\frac{2}{3}) + 2 \times
({\bf \overline{4}},{\bf 22},{\bf 1},{\bf 1};
\frac{1}{3}, -\frac{1}{3},-\frac{1}{3})$ & \\
& $ ~2 \times
({\bf 6},{\bf 1},{\bf 1},{\bf 1};
\frac{4}{3}, -\frac{1}{3},\frac{2}{3}) + 2 \times
({\bf \overline{4}},{\bf 1},{\bf 3} \hbox{ and }
{ \bf \overline{3}}, {\bf 1}; \frac{1}{3},\frac{2}{3},-\frac{1}{3})$
& $(1,\alpha,\alpha, \alpha)$ \\
&
$~2 \times
({\bf \overline{4}},{\bf 1},{\bf 1},{\bf 4}_S \hbox{ and }
{\bf 4}_{S'}; \frac{1}{3},\frac{2}{3},\frac{1}{6}) + 2 \times
({\bf 1},{\bf 1},{\bf 3} \hbox{ and }
{\bf \overline{3}},{\bf 6}; -\frac{2}{3},-\frac{1}{3},-\frac{1}{3})$ & \\
 & $ ~2 \times
({\bf 4},{\bf 1},{\bf 1},{\bf 1};
-\frac{5}{3},\frac{2}{3},-\frac{1}{3}) + 2 \times
({\bf 1},{\bf 1},{\bf 1},{\bf 1};
-\frac{2}{3},-\frac{1}{3},-\frac{4}{3})$ & \\
& & \\ \hline
& & \\
total & & $(\alpha,\alpha,\alpha,\alpha)$  \\
& & \\ \hline
\end{tabular}
\caption{Nonsupersymmetric asymmetric $Z_6$ model with  $\beta_L=
(\frac{1}{6}^4,\frac{1}{3},0^{11};0^2;\frac{1}{3},0^3)$, 
$\alpha^3=1$.}}

\TABLE[h]{
\begin{tabular}{|c|c|c|} \hline
sector & $~SU(3)\times SO(26)\times SU(3)'\times
SO(8) \times U(1) $ & $Z_3$ anomaly \\ \hline
& & \\
untwisted   & $ ~ 2 \times ({\bf 3},{\bf 26},{\bf 1},{\bf 1}; 1;r_1) +
2 \times ({\bf 3},{\bf 1},{\bf 1},{\bf 1}; -2;r_1)$
& $(1,1,1,1)$
\\
&  & \\ \hline
$n=1$  & ~ {no states} & \\ \hline
$n=2$ &  ~ no states  & \\ \hline
& & \\
$ n=3 $ &  $~2 \times ({\bf \overline{3}},{\bf 1},{\bf 1},
{\bf 8}_V \hbox{ and }
{\bf 8}_S \hbox{ and } {\bf 8}_{S'};-1) $ & $(1,1,1,1)$  \\
& & \\ \hline
& & \\
$n=4$ &   $~2 \times ({\bf 1},{\bf 26},{\bf 3}
\hbox{ and } {\bf \overline{3}},{\bf 1};-1) +
2 \times ({\bf 1},{\bf 1},{\bf 3}
\hbox{ and } {\bf \overline{3}},{\bf 1};2) $ & $(1,1,1,1)$  \\
 & $~2 \times ({\bf \overline{3}},{\bf 1},{\bf 1},{\bf 1};-2) +
2 \times ({\bf \overline{3}},{\bf 26},{\bf 1},{\bf 1};1)$ & \\
& & \\ \hline
$n=5$ & $~2 \times ({\bf 3},{\bf 1},{\bf 1},{\bf 8}_V
\hbox{ and } {\bf 8}_S \hbox{ and } {\bf 8}_{S'};-1)$ & $(1,1,1,1)$ \\
& $~2 \times ({\bf 1},{\bf 1},{\bf 3} \hbox{ and }
{\bf \overline{3}},{\bf 8}_V \hbox{ and } {\bf 8}_S \hbox{ and }
{\bf 8}_{S'};1) $ & \\
& & \\ \hline
\end{tabular}
\caption{Nonsupersymmetric
asymmetric $Z_6$ model with $\beta_L=(\frac{1}{3}^3 ,0^{13};0^2;0^4)$.}}

\subsection{Supersymmetric Examples}

Let us turn, now, to supersymmetric theories.

\subsubsection{Narain--Sarmadi--Vafa}
 
It is instructive to first
consider the Narain--Sarmadi-Vafa (NSV) supersymmetric
asymmetric model \cite{ao}. The internal lattice is a product of the
$E_8 \times E_8$ lattice with the
lattices $(\Gamma_{(2,2)}(A_2) )^3$.
As discussed before the $\Gamma_{(2,2)}(A_2)$
lattice has independent left and right
$Z_3$ symmetries, and a left--right symmetric
$Z_6$ symmetry.
The NSV model corresponds to performing an
asymmetric $Z_3$ twist
\beq
\beta_L = \phi_R = {1 \over 3} (-2,1,1) ~,
\eeq
with no left-moving twist or right-moving
shift ($\phi_L=0$ and
$\beta_R=0$).
This is modular invariant, and preserves $N=1$ supersymmetry.
The low-energy gauge group is $E_8 \times E_6 \times SU(3)^4$.
The gauge twist breaks one $E_8$ group and accounts for
one of the $SU(3)$ group factors.
The other three
$SU(3)$ group factors correspond to
left-moving lattice
momentum in the three $\Gamma_{(2,2)}(A_2)$ lattices, which
would otherwise be projected out in a symmetric orbifold
construction of the same lattice.

The spectrum is easily worked out. The massless fermions
in the untwisted
sector are the gauginos, and from the matter
$3 \times ({\bf 3}, {\bf 27}; {\bf 1}, {\bf 1}, {\bf 1})$.
In the twisted sector we have  \cite{ao}
\ba
& ({\bf 3}, \overline{\bf 27}; {\bf 1}, {\bf 1}, {\bf 1})
+\left(({\bf 1}, {\bf 27}; {\bf 3} + \overline{\bf 3}, {\bf 1}, {\bf 1})
+~\hbox{permutations in last three groups} \right) \nonumber \\
& + \left((\overline{\bf 3}, {\bf 1}; {\bf 3}+ \overline{\bf 3},
{\bf 1}, {\bf 3} + \overline{\bf 3})+ ~\hbox{permutations in last 3
groups} \right) ~.
\ea

Focusing on one of the $\Gamma_{(2,2)}(A_2)$ lattices,
we can look for
possible discrete anomalies in the
$Z_3$ symmetry that acts only on the right-movers in
this lattice.
In the untwisted (RM) Ramond sector, the positive
helicity massless states are
$({1 \over 2}, {1 \over 2}, \pm {1 \over 2}, \pm {1 \over 2})$,
$({1 \over 2}, -{1 \over 2}, \pm {1 \over 2}, \mp{1 \over 2})$.
Here as before we are using the bosonized formulation.
In this case the gauginos are obtained from the
state with all `$+$'s and have $Z_3$ charge
$\gamma^{-1/2}$, whereas the matter states are obtained from
the three remaining right-movers;  two have charge $\gamma^{+1/2}$
and
the other has charge $\gamma^{-1/2}$.
The $(SU(3), E_6, SU(3)_{I=1,2,3})$ discrete
anomalies in the untwisted sector are
$(\gamma^{3/2},1,1,1,1)$,
where it is important to include the contribution from
the gauginos. It is useful to recall that
$t_2({\bf 27})=6$ and $t_2({\bf 78})=24$ in the normalization of
Table \ref{zeromodes}. We
will find that this
anomaly is canceled by a contribution from the twisted sector states.

In the twisted sector the
discrete charges of the
massless states receive two contributions. The first is from
the twisted fermions. The Ramond RM ground state is unique and
massless,
and corresponds to the state $r + \phi$, with
$r=({1 \over 2}, {1 \over 2},
-{1 \over 2}, - {1 \over 2})$.
Using the fermionic twist operator, the world-sheet
fermion contribution to the discrete
charge of the twisted spacetime fermions is easily found to
be $\gamma^{1/6}$. The twisted bosons provide a
charge that is either $\gamma^{1/3}$ or $\gamma^{-2/3}$. Since for any
twisted state the number
of zero modes in a one instanton background is a multiple
of 9, the bosonic charge does not contribute to an anomaly.
In total the discrete anomalies from the twisted states are
$(\gamma^{3/2}, 1,1,1)$, where the previous notation has
been used. Although this is non-universal,
a cancellation occurs
between the twisted and untwisted
sectors, leaving no $Z_3$ anomaly.

\subsubsection{$Z_6$ models}

Now a class of supersymmetric models with orbifold group
$Z_6$ is constructed. We have already seen
that the hexagonal lattice $\Gamma_{(2,2)}(A_2)$
has separate left and right $Z_3$ symmetries.
But the hexagonal lattice
also has a $Z_6$ automorphism, which
can be used to construct symmetric orbifold twists.
Since this rotation exchanges the two $SU(3)$ fundamental weights,
it cannot be used to construct an asymmetric twist.
But a 
symmetric $Z_6$ twist in this plane is allowed,
since this is a symmetry of the lattice.
(The symmetric $Z_6$ twist
preserves the condition $p_L - p_R \in R$, whereas the asymmetric
twist does not.)

The lattice we therefore consider the product 
of the internal $SO(32)$ and 
the  $(\Gamma_{(2,2)}(A_2))^3$ lattices.
We orbifold by an asymmetric
$Z_3$ twist on the first $SU(3)$ lattice, and a symmetric
$Z_6$ twist on the last two $SU(3)$ lattices.
In particular, consider
\beq
\phi_R=(-1/3,1/6,1/6)  ~~; ~~
\phi_L=(0,1/6,1/6) ~.
\eeq
This
preserves $N=1$ supersymmetry.
It is straightforward to see that the last (non-trivial) 
level matching condition (\ref{lm2})is
satisfied. The details may be found in Appendix B. 
In addition, different models will be characterized
by a shift $\beta_L$ on the left--movers.
These must satisfy the level matching conditions (\ref{lm1}).

\TABLE[h]{
\begin{tabular}{|c|c|} \hline
 sector & Ramond state  \\ \hline
& \\
untwisted   &
$r_0=(\frac{1}{2},  \frac{1}{2},  \frac{1}{2},  \frac{1}{2}
; \alpha^{- 1/2}) ~,~r_2=
(\frac{1}{2},  \frac{1}{2}, - \frac{1}{2}, - \frac{1}{2}
; \alpha^{- 1/2})$ \\
 & $r^{\pm}_3=({1 \over 2}, - {1 \over 2}, \pm {1 \over 2}, \mp {1 \over 2} ;
\alpha^{1/2})$ \\
& \\  \hline
& \\
$n=1$  &   $~(\frac{1}{2},\frac{1}{2}, - \frac{1}{2}, -\frac{1}{2};
\alpha ^{1/2} ~ \hbox{if} ~ p_{(1),L} \in Y_0   ~
\hbox{or} ~ \alpha ^{-1/2} ~ \hbox{if} ~ p_{(1),L} \in Y_{1,2} ) $ \\
& \\  \hline
& \\
$n=2$ & $~(\frac{1}{2},\frac{1}{2}, - \frac{1}{2},
-\frac{1}{2}
; \alpha ^{-1/2} ~ \hbox{if} ~ p_{(1),L} \in Y_0 ~
\hbox{or} ~ \alpha ^{1/2} ~ \hbox{if} ~ p_{(1),L} \in Y_{1,2}
 ) $ \\
& \\ \hline
& \\
$ n=3 $ & $~(\frac{1}{2},\frac{1}{2}, - \frac{1}{2}, -\frac{1}{2};
\alpha^{1/2} ) $  \\
& \\ \hline
\end{tabular}
\label{susytable}
\caption{Supersymmetric asymmetric $Z_6$ model: Positive helicity
RM supersymmetric Ramond sector weights $(r)$ and the
Ramond sector
$Z_3$ charges $(\alpha^3=1)$. The ground states are $r+n \phi_R$.}}

Proceeding as
before, the projection operators for massless states in the
$nth$-twisted sectors are:
\ba
n=0: & ~ & p _L\cdot \beta_L  - r \cdot \phi_R + N_{osc} =~0 ~\hbox{mod
1}
 \\
n=1: & ~ & p_L \cdot \beta_L  - r \cdot \phi_R + \frac{1}{2}(\beta^2 _L
-
\phi^2_L)
+ \frac{1}{2} \phi^{1} _R  \nonumber \\
& & + \left \{ ~{1 \over 3} ~\hbox{if} ~
p_{(1),L} \in Y^{(1)} _{i=1,2} ~\hbox{or} ~ 0 ~\hbox{if}
~ p_{(1),L} \in Y_0 \right \} +N_{osc} ~= ~0 ~\hbox{mod 1}
\label{projn1} \\
n=2: & ~ &   p_L \cdot \beta_L  - r \cdot \phi_R + (\beta^2 _L-
\phi^2_L)
+ \frac{1}{2} \phi^{1} _R + \frac{1}{2} \nonumber \\
& & + \left \{~{ 2 \over 3}
~\hbox{if}
~ p_{(1),L} \in Y^{(1)} _{i=1,2} ~\hbox{or} ~ 0 ~\hbox{if}
~ p_{(1),L} \in Y_0 \right \} +N_{osc} \nonumber \\
& & =~ \{ ~0 ~\hbox{mod
1}
~ \hbox{or} ~ {1 \over 2} ~\hbox{mod 1} \}
\label{projn2}
\\
n=3: & ~ & p_L \cdot \beta_L - r \cdot \phi_R + {3 \over 2} (\beta^2 _L
- \phi^2_R) + \phi^1 _R + {1 \over 2} + N_{osc} \nonumber \\
& & = ~\{
0 ~\hbox{mod 1} ~ \hbox{or} ~ {1 \over 3}  ~\hbox{mod 1 or } ~ {2 \over
3} ~\hbox{mod 1} \} ~,~~{p_{(1),L} \in Y_0} ~.
\label{projn3}
\ea
Here $(p_{(1),L},p_{(1),R})$ are momenta in the
lattice with the asymmetric twist.
In addition, the bosonic degeneracy
for each choice of phase appearing on the right-side of the above
equations is required.
In the
singly twisted sector the bosonic degeneracy is 1.
For
$n=2$, the projector is
\beq
P = \frac{1}{6}(9 + \Delta + 9 \Delta ^2
+ \Delta ^3  + 9 \Delta^4 +
\Delta ^5) ~,
\eeq
where $\Delta= e^{ 2 \pi i \Phi}$  and $\Phi$ is given
by the expression appearing on the left side
of
(\ref{projn2}). Thus the degeneracy of states with $\Phi= 0$ mod 1 is
5, whereas the degeneracy of states with $\Phi=1/2$ mod 1 is 4.
Similarly, in the $n=3$ sector the projector is
\beq
P = \frac{1}{6}(16+  \Delta +  \Delta ^2
+ 16 \Delta ^3  + \Delta^4 +
\Delta ^5).
\eeq
Again, $\Delta$ is defined as before with now $\Phi $ equal
to the expression appearing on the
left side of (\ref{projn3}). The degeneracy of states with
$\Phi =0$ mod 1 $(\Delta=1)$
is 6, and those with $\Phi =1/3$ mod 1 $(\Delta=\gamma^2)$
or
$2/3$ mod 1 $(\Delta =\gamma^4)$ is 5.

There are a number of discrete anomalies that may be studied.
There is a $Z_3$ symmetry of the $\Gamma_{(2,2)}(A_2)$
lattice with the asymmetric twist. This is the same
asymmetric $Z_3$ symmetry studied in the non-supersymmetric
models. The discrete charges are computed as in the
previous section and listed in Table \ref{susytable}.

There is also a symmetric
$Z_6$ symmetry acting on either of the $\Gamma_{(2,2)}(A_2)$
lattices with the symmetric twist. Here now
the $Z_6$ charges
in the twisted sector
also depend on the fixed point. This too has been 
described in the section on symmetric orbifolds.

In addition there is another symmetric $Z_6$ symmetry
acting on the $\Gamma_{(2,2)}(A_2)$ lattice
that has the asymmetric twist. But since this lattice
has an enhanced $SU(3)$ gauge symmetry,
the $Z_6$
symmetry permutes, for example, the ${\bf 3}$ and the
$\overline{\bf{3}}$ of $SU(3)$. It is unclear how to study the
instanton anomalies of this symmetry.
Finally there is the quantum $Z_N$ symmetry.

The massless spectra for three models is given in
Tables \ref{susyI}, \ref{susyII} and \ref{susyIII}.
Again, perturbative anomalies cancel, with a Green--Schwarz
mechanism required for models II and III. This provides a
non-trivial consistency check on the models.
Discrete anomalies are computed as in the
previous section.  Of course, one must now take
account of the gaugino contribution from the untwisted sector.
The quantum
$Z_6$ anomalies cancel as well, although a discrete ``Green--Schwarz''
mechanism is required for models II and III. The symmetric
$Z_6$ symmetry is also found to be non-anomalous, again,
with a discrete ``Green--Schwarz mechanism'' required for models II and III.
The asymmetric $Z_3$ anomalies also cancel, but with a discrete
``Green--Schwarz mechanism'' required for model III.
It is interesting to note that in all of these
models the
discrete anomalies are non-universal in the
untwisted sector, and only become universal
after adding in the contributions from the
twisted sectors.

\TABLE[h]{
\begin{tabular}{|c|c|c|c|} \hline
sector & $SU(2) \times SO(28) \times SU(3) \times U(1) $
& $Z_3$ anomaly & $Z_6$ anomaly  \\ \hline
& & & \\
untwisted   & $~({\bf 3},{\bf 1},{\bf 1}; 0;r_0)
+ ({\bf 1},{\bf 373},{\bf 1};0;r_0)
+({\bf 1},{\bf 1},{\bf 8}; 0;r_0) $ & $(\alpha^2,\alpha^2,1)$ &
$(\gamma^2,\gamma^2,1)$ \\
& & & \\
& $  ({\bf 2},{\bf 28},{\bf 1}; 1;r^{\pm}_3) $
& &   \\
& & & \\ \hline
& & & \\
$n=1$  & $~({\bf 2},{\bf 28},{\bf 1};-\frac{2}{3})
+ 4 \times ({\bf 1},{\bf 1},{\bf 3} \hbox{ and }
{\bf \overline{3}}; \frac{1}{3})$  &
$(\alpha^2,\alpha^2,\alpha^2)$ &
$(\gamma^{2/3},\gamma^{8/3},
\gamma^{-2/3})$
\\
& & & \\ \hline
& & & \\
$n=2 $ & $ ~4 \times
({\bf 2},{\bf 28},{\bf 1};- \frac{1}{3}) +
5 \times ({\bf 1},{\bf 1},{\bf 3}
\hbox{ and } {\bf \overline{3}};-\frac{4}{3})$
& $(\alpha,\alpha,\alpha)$ & $(\gamma^{4/3},\gamma^{-2/3},
\gamma^{2/3})$ \\
& & & \\
 &
$~8 \times ({\bf 1},{\bf 1},{\bf 3}
\hbox{ and }  {\bf \overline{3}}; \frac{2}{3})$ & &   \\
& & & \\ \hline
& & & \\
$ n=3 $ & $~5 \times
({\bf 2},{\bf 28},{\bf 1}; 0) $ & $(\alpha,\alpha,1)$ &
$(\gamma^{2},\gamma^2,1)$ \\
& & &  \\ \hline
& & & \\
total & & $(1,1,1)$ & $(1,1,1)$ \\
& & & \\ \hline
\end{tabular}
\caption{Supersymmetric asymmetric $Z_6$ 
model I: $\beta_L=({1 \over 6} ^2, 0^{14};
0^2)$, $\alpha^3=1$ and $\gamma^6=1$. Some states have 
oscillators.}
\label{susyI}}

\footnotesize
\TABLE[h]{
\begin{tabular}{|c|c|c|c|} \hline
sector & $ SU(2) \times SU(6)  \times SO(16) \times SU(3) \times U(1)^2 $
& $Z_3$ anomaly & $Z_6$ anomaly \\ \hline
& &  & \\
untwisted & $ ({\bf 3}, {\bf 1}, {\bf 1}, {\bf 1};0,0;r_0)
~+~ ({\bf 1},{\bf 35}, {\bf 1}, {\bf 1};0,0;r_0) $
& $(\alpha^2,\alpha,\alpha^2,1)$ & $(\gamma^2,\gamma^2,\gamma^2,1)$ \\
& & & \\
& $ ({\bf 1}, {\bf 1}, {\bf 120}, {\bf 1};0,0;r_0) ~+~
 ({\bf 1}, {\bf 1},{\bf 1}, {\bf 8};0,0;r_0) $ & & \\
& & & \\
& $  ({\bf 2}, {\bf 1}, {\bf 16}, {\bf 1};1,0;r^{\pm}_3)
~+~ ({\bf 1}, {\bf 6}, {\bf 16}, {\bf 1};0,-1;r_2) $ & & \\
& & & \\
& $ ({\bf 2}, \overline{\bf 6}, {\bf 1}, {\bf 1};-1,1;r^{\pm}_3)
~+~ ({\bf 1}, \overline{\bf 15}, {\bf 1}, {\bf 1};0,2;r_2) $ & & \\
& & & \\ \hline
& & &  \\
$n=1$ & $({\bf 2}, {\bf 6}, {\bf 1}, {\bf 1};-{2 \over 3}, 1)
~+~ 2 \times ({\bf 1}, {\bf 1}, {\bf 1}, {\bf 3} ~\hbox{and}
~ \overline{\bf 3};{1 \over 3},2) $ &
$(1,\alpha^2,1,\alpha)$ & $(\gamma^4,\gamma^{8/3},1,\gamma^{2/3})$ \\
& & & \\
& $2 \times ({\bf 1}, {\bf 15}, {\bf 1}, {\bf 1};{1 \over 3}, 0)
$ & & \\
& & & \\ \hline
& & & \\
$n=2$ & $ 5 \times ({\bf 1}, {\bf 1}, {\bf 1}, {\bf 3} ~\hbox{and}
~ \overline{\bf 3};{2 \over 3},-2) ~+~ 4 \times
({\bf 2}, \overline{\bf 6}, {\bf 1}, {\bf 1};-{1 \over 3}, -1) $
& $(1,\alpha^{-2},1,\alpha^2)$ & $(\gamma^2, \gamma^{-8 /3},1,
\gamma^{10/3})$  \\
& & & \\
& $ 5 \times ({\bf 1}, \overline{\bf 15}, {\bf 1}, {\bf 1};{2 \over 3},0)
$ & & \\
& & & \\ \hline
& & & \\
$n=3$ &  $5 \times ({\bf 2}, {\bf 1}, {\bf 16}, {\bf 1};0,0)
~+~ 5 \times ( {\bf 2}, \overline{\bf 6}, {\bf 1}, {\bf 1};0,1) $
& $(\alpha,\alpha^2,\alpha,1)$ & $(\gamma^2,\gamma^2,\gamma^2,1)$ \\
&  &  & \\
& $ 6 \times ({\bf 2}, {\bf 6}, {\bf 1}, {\bf 1};0,-1)$ & & \\
& & & \\ \hline
$n=4$ & none & & \\ \hline
$n=5$ & none & & \\ \hline
& & & \\
total & & $(1,1,1,1)$ & $(\gamma^4,\gamma^4,\gamma^4,\gamma^4)$  \\
& & & \\ \hline
\end{tabular}
\caption{Supersymmetric asymmetric model II: $\beta_L=({1 \over 6}^2,{1 \over 3}^6,
0^8;0^2;0^4)$, $\alpha^3=1$ and $\gamma^6=1$.}
\label{susyII}}

\TABLE[h]{
\begin{tabular}{|c|c|c|c|} \hline
sector & $~SU(5) \times SU(3) \times SO(14) \times
SU(3) \times U(1)^3 $ & $Z_3$ anomaly & $Z_6$ anomaly \\ \hline
& & & \\
 & $~({\bf 24}, {\bf 1},{\bf 1},{\bf 1};0,0,0;r_0)
+({\bf 1}, {\bf 8},{\bf 1},{\bf 1};0,0,0;r_0) $ & & \\
& & & \\
&
$({\bf 1}, {\bf 1}, {\bf 91}, {\bf 1};0,0,0;r_0) +
({\bf 1},{\bf  1},{\bf  1} ,{\bf  8}; 0,0,0;r_0) $ & & \\
& & & \\
untwisted & $~({\bf 1},{\bf \overline{3}},{\bf 14},{\bf 1}; 0,-1,0;r_2)
+({\bf 1},{\bf \overline{3}},{\bf 1},{\bf 1};0,2,0;r_2) $  &
 $(\alpha^{1 \over 2},\alpha^{5 \over 2}, \alpha, 1 )$ &
$(\gamma, \gamma^3,1,1)$ \\
& & & \\
& $ ~({\bf \overline{10}},{\bf 1},{\bf 1},{\bf 1}; -2,0,0;r_2)
+({\bf 5},{\bf 1},{\bf 1},{\bf 1}; 1,0,\pm 1;r_2)$ &  & \\
& & & \\
& $ ({\bf 5},{\bf 1},{\bf 14},{\bf 1}; 1,0,0;r^{\pm}_3)
+  ({\bf \overline{5}},{\bf 3},{\bf 1},{\bf 1};-1,1,0;r^{\pm}_3) $
& & \\
& & & \\
& $ 
({\bf 1},{\bf \overline{3}},{\bf 1},{\bf 1}; 0,-1,\pm 1;r^{\pm}_3)
 $  & & \\
&  & & \\ \hline
&  & & \\
 &  $~({\bf \overline{5}},{\bf  \overline{ 3}},{\bf 1}
,{\bf 1};-{1 \over 6} ,0,{1 \over 2})+
({\bf 1},{\bf \overline{3}},{\bf 1},{\bf  3}
 \hbox{ and } {\bf \overline{3}};{5 \over 6},0,-{1 \over 2})
$ & & \\
& & & \\
 $n=1 $ & $~ ({\bf 1},{\bf 1},{\bf 14},{\bf 1};{5 \over 6},1,-{1 \over 2})
+2 \times ({\bf \overline{5}},{\bf 1},{\bf 1},{\bf 1}; -{1 \over 6},
1,-{1 \over 2}) $ & $(\alpha^{5 \over 2}, \alpha^2, \alpha, \alpha)$
& $(\gamma^{-{2 \over 3}}, \gamma^{2 \over 3}. \gamma^{4 \over 3},
\gamma^{-{4 \over 3}})$ \\
& & & \\
 & $~2 \times ({\bf 1},{\bf 3},{\bf 1},{\bf 1};{5 \over 6},-1,
{ 1 \over 2}) +
2 \times ({\bf 1},{\bf 1},{\bf 1}, {\bf 3} \hbox{ and }
{\bf \overline{3}};{5 \over 6},1,{1 \over 2})$ & & \\
& & & \\
&  $ 3 \times
({\bf 1},{\bf \overline{3}},{\bf 1},{\bf 1};{5 \over 6},0,
-{1 \over 2} ) $ & & \\
& & & \\ \hline
& & & \\
$n=2$ & $~ 4 \times ({\bf \overline{10}},{\bf 1},{\bf 1},{\bf 1}
;-{1 \over 3},-1,0) + 4 \times
({\bf 1},{\bf \overline{3}},{\bf 1},{\bf 1};{5 \over 3},1,0) $
& $(\alpha^{3 \over 2}, \alpha^{1 \over 2}, 1, \alpha )$
& $(\gamma^3, \gamma^{-{7 \over 3}}, 1, \gamma^{8 \over 3})$ \\
& & & \\
& $~4 \times
({\bf 1},{\bf 1},{\bf 1}, {\bf 3} \hbox{ and } {\bf \overline{3}};
{5 \over 3},-1,0) + 5 \times
({\bf \overline{5}},{\bf 3},{\bf 1},{\bf 1};{2 \over 3},0,0)$
& & \\
& & & \\ \hline
& & & \\
$ n=3 $ & $~5 \times ({\bf 5},{\bf 1},{\bf 1},{\bf 1};
-{3 \over 2},0,{1 \over 2}) + 5 \times
({\bf \overline{5}},{\bf 1},{\bf 1},{\bf 1}; {3 \over 2},0,
-{1 \over 2})$ & $(\alpha^{1 \over 2}, 1, 1,1)$
& $(\gamma^4, 1,1,1)$ \\
& & & \\
& $~5 \times ({\bf \overline{10}},{\bf 1},{\bf 1},{\bf 1};
{1 \over 2},0,{1 \over 2}) + 6 \times
({\bf 10},{\bf 1},{\bf 1},{\bf 1}; -{1 \over 2},0,-{1 \over 2}) $
& & \\
& & & \\ \hline
& & & \\
total & & $(\alpha^2, \alpha^2, \alpha^2, \alpha^2)$
& $(\gamma^{4 \over 3}, \gamma^{4 \over 3}, \gamma^{4 \over 3},
\gamma^{4 \over 3})$ \\
& & & \\ \hline
\end{tabular}
\caption{Supersymmetric asymmetric model III: $\beta_L=({1 \over 6}^5,{1 \over 3}^3,
{1 \over 2}, 0^7;0^2)$, $\alpha^3=1$ and $\gamma^6=1$.}
\label{susyIII}}
\normalsize

\subsubsection{Universal Couplings of Moduli}
\label{othersusymoduli}

One must again ask whether other moduli might cancel the
anomalies.  In this case, the question is not quite as simple as
in the non-supersymmetric examples of the previous subsection.
There are now moduli, neutral under all symmetries, in the
untwisted sector.  However, using the methods of
\cite{kaplunovsky}, one can readily show that these moduli couple
universally to all gauge groups, in a fashion similar to the
weakly coupled string dilaton. 
As a
result, they could not have helped with anomaly cancellation.

\section{A Search for Discrete Gauge Anomalies in Type IIB
Orientifolds}
\label{typeiibsec}

This section describes models obtained by the  
orientifold compactification of 
Type IIB string theory on $T^6$. At low-energies 
they describe four--dimensional,  
${\cal N}=1$ supersymmetric theories. 
In contrast to the heterotic models 
discussed above, the Type IIB orientifolds models 
discussed here {\em do} have non--universal discrete anomalies. 
If these are to be 
canceled there must be massless states with non--universal 
couplings to the gauge bosons. But how is this possible?  
For the only
untwisted scalars that are not charged under the $Z_M$ discrete
symmetry 
are the dilaton multiplet and the $T$
moduli. In the first instance the coupling to D9 brane
matter is universal, and in the second the $T$ moduli
do not couple to D9 matter at tree-level. 

If this were the end of the story then there 
would be a puzzle. 
But there is an important
distinction between orientifolds and orbifolds.
In orientifold models the
world--sheet parity projection breaks the quantum $Z_N$ symmetry.
Consequently, a linear coupling of a 
massless R-R state in the twisted sector to gauge bosons 
is not forbidden. 
Non-universal
discrete anomalies
can then be canceled by
assigning a shift to
these moduli.  Indeed this is not surprising, 
for such couplings are needed
to cancel the multiple $U(1)$ anomalies typically
found in these class 
of models \cite{ibanezdbrane1,ibanezdbrane2}.

\subsection{Discrete Charges}

The symmetries we study are the discrete isometries
of the internal manifold. Initially there are 
three such isometries, but the orientifold 
projection removes one. 
We initially 
focus on the $Z_M$ symmetry of the third torus,
which
in our notation corresponds to the last entry in $r$.
This symmetry acts only
on the world sheet variables and not the Chan-Paton
factors.
The charges of the states can be
directly read off from (\ref{99matter}), (\ref{95matterb})
and (\ref{95matterf}): 

$\bullet$ The $Z_M$ charge of a 99 or 55 state (\ref{99matter})
with $r=(\cdots \pm)$ is simply
$\gamma^{\pm/2}$, where $\gamma^M=1$. Clearly this is 
a discrete $R$--symmetry.

$\bullet$ In the
95 sector the spacetime bosons are neutral under
this $Z_M$, whereas spacetime fermions with
negative helicity $(s_0=-1/2)$ have charge
$\gamma^{1/2}$.

In models where the orbifold group 
has a single $Z_N$ factor
one may also study the remaining 
discrete rotation, which is  
the product of a discrete $Z_6$ rotation in the first torus
with an anti--rotation in the third torus.
The gauginos are neutral
under this linear combination, so this is a non--$R$ discrete 
symmetry. 
The charges of the other states are easily obtained from
the formulae given above.

\subsection{Models}

The three models we study have been constructed and
presented in the literature \cite{ibanezdbrane1,ibanezdbrane2}.
The new ingredient here is the computation of
the discrete anomalies.

\subsubsection {$Z_3$}

Here the twist is generated
by $\phi=(1,1,-2)/3$. This model has only D9 branes.
The embedding of the twist into
the gauge group is described by the shift vector
$V=(1^{12},0^4)/3$. The gauge group is
$U(12) \times SO(8)$. The massless charged fermions and gauginos
are \cite{ibanezdbrane1} (for $s_0=-1/2$)
\ba
99 & : & 2({\bf 12},{\bf 8};1;\gamma^{1/2}) +
({\bf 12},{\bf 8};1;\gamma^{-1/2}) +
2 (\overline{\bf 66}, {\bf 1};-2,\gamma^{1/2}) + \\ \nonumber
& & (\overline{\bf 66}, {\bf 1};-2,\gamma^{-1/2})
+({\bf 143},{\bf 1};0,\gamma^{-1/2})
+({\bf 1}, {\bf 28};0,\gamma^{-1/2})
\ea
with the second-to-last entry indicating the $U(1)$ charge
and the last entry indicating the $Z_6$ charge.
The $U(1)$ in this model is anomalous and non-universal.
In particular, the mixed $U(1)$
and non-abelian gauge anomalies are :
$(U(1) SU(12)^2, U(1) SO(8)^2) \propto (1,-2)$ \cite{ibanezdbrane2}.
This can be canceled by a shift in one of the neutral twisted
R-R scalars \cite{ibanezdbrane2}.

Now we compute the anomalies for the $Z_6$ symmetry
of the third torus, say. The $Z_6$ charges of
the matter and gauginos are easily computed using
the rule given in the previous section. One finds
non-universal discrete anomalies
\beq
(Z_6 SU(12)^2,Z_6 SO(8)^2) =(\gamma^3,1)~.
\label{dbranemodel1dis}
\eeq
Note that this is non-universal. But
since the discrete anomalies are in the
same ratio as the $U(1)$ anomalies, they
may be canceled by assigning a shift to
the R-R scalar ${\cal M}$ used to cancel the $U(1)$ anomalies.
{\it In fact, one can define discrete symmetries which are unbroken
and free of anomalies by combining the discrete transformations
with suitable $U(1)$ transformations.}  

One can also consider the non--$R$ $Z_6$ symmetry which is a
combination of a
rotation in the third torus with an anti-rotation in, say,
the first torus. 
One finds that this symmetry is not anomalous.

\subsubsection {$Z_3 \times Z_3$}

Here the first $Z_3$ is generated by a
twist with $\phi_1=(1,-1,0)/3$ and the second
by a twist with $\phi_2=(0,1,-1)/3$.
The shifts associated with these twists
are $V_1=(1^4,-1^4,0^8)/3$ and
$V_2=(0^4,1^4,-1^4,0^4)/3$. The gauge group
is then $U(4)^3 \times SO(8) $.
The massless charged fermions and gauginos
are \cite{ibanezdbrane1} (for $s_0=1/2$)
\ba
99 & : &
\nonumber
({\bf 4},{\bf 4},{\bf 1}, {\bf 1};1,1,0;\gamma^{1/2})
+
({\bf 4},{\bf 1}, {\bf 4}, {\bf 1};1,0,1;\gamma^{-1/2})
 +
({\bf 1},{\bf 4}, {\bf 4}, {\bf 1};0,1,1;\gamma^{-1/2})+
 \\
 \nonumber
& &
({\bf 6},{\bf 1},{\bf 1},{\bf 1};2,0,0;\gamma^{-1/2})
+
({\bf 1}, {\bf 6},{\bf 1},{\bf 1};0,2,0;\gamma^{-1/2})
 +  ({\bf 1},{\bf 1}, {\bf 6},{\bf 1};0,0,2;\gamma^{1/2})+ \\
  \nonumber
& & (\overline{\bf 4},{\bf 1}, {\bf 1};{\bf 8};-1,0,0;\gamma^{-1/2})
+
({\bf 1}, \overline{\bf 4}, {\bf 1};{\bf 8};0,-1,0;\gamma^{-1/2})
 +  ({\bf 1},{\bf 1},  \overline{\bf 4};{\bf 8};0,0,-1;\gamma^{1/2})+
\\
 \nonumber
& & ({\bf 15}, {\bf 1}, {\bf 1}, {\bf 1};0,0,0;\gamma^{1/2})
+
({\bf 1}, {\bf 15}, {\bf 1}, {\bf 1};0,0,0;\gamma^{1/2})
 +  ({\bf 1}, {\bf 1}, {\bf 15}, {\bf 1};0,0,0;\gamma^{1/2})+
\\
& &  ({\bf 1}, {\bf 1}, {\bf 1}, {\bf 28};0,0,0;\gamma^{1/2})
\label{dbranemodel2}
\ea
The $U(1)$'s in this model are anomalous and given by
\cite{ibanezdbrane2} $
(SU(4)^2_i U(1)_j; SO(8)^2 U(1)_j) \propto (1,1,1;-2) $ .
The $Z_6$ charges of
the matter and gauginos are easily computed using
the rule given in the previous section and are
given by
the last entry of each item in (\ref{dbranemodel2}).
One finds the
discrete anomalies
\beq
(Z_6 SU(4)^2_i;Z_6 SO(8)^2) =(\gamma^{-1},\gamma^{-1},\gamma^{-1};\gamma^2)~,
\eeq
where $\gamma^6=1$.
As in the previous model
these are non-universal, but 
in the
same ratio as the $U(1)$ anomalies. 
Thus the twisted R-R scalar that is responsible
for canceling the $U(1)$ anomalies may also be used
to cancel the discrete anomalies.  Alternatively,
we may again define a combination of the original
discrete symmetry transformation and discrete
$U(1)$ transformations to define unbroken anomaly
free discrete symmetries.

\subsubsection {$Z_6$}

This orbifold is generated by the $Z_6$
twist $\phi=(1,1,-2)/6$. Since this group contains
an element of order 2, this model has both D5 and D9
branes.
For simplicity the D5 branes are placed at
the origin. Then with the choice of shifts 
$V_{99}=V_{55}=(1^6,5^6,3^4)/12$ that satisfy 
the tadpole cancellation requirement, the gauge group
is $U(6)\times U(6) \times U(4)$ for the 99 matter
and a different 
$U(6)\times U(6) \times U(4)$ for the 55 matter.
The massless charged fermions and gauginos are  \cite{ibanezdbrane1}
(for $s_0=-1/2$)
\ba
99 & : & 2({\bf 15},{\bf 1},{\bf 1};{\bf 1},{\bf 1},{\bf 1};\gamma^{1/2})
+({\bf 6}, \overline{\bf 6},{\bf 1};{\bf 1},{\bf 1},{\bf 1};\gamma^{-1/2}) +
(\overline{\bf 6},{\bf 1},\overline{\bf 4};{\bf 1},{\bf 1},{\bf 1};
\gamma^{-1/2}) +
\nonumber \\
& &
2(\overline{\bf 6},{\bf 1},{\bf 4};{\bf 1},{\bf 1},{\bf 1};\gamma^{1/2})+
({\bf 1},{\bf 6},{\bf 4};{\bf 1},{\bf 1},{\bf 1};\gamma^{-1/2}) +
2({\bf 1},{\bf 6},\overline{\bf 4};{\bf 1},{\bf 1},{\bf 1};\gamma^{1/2}) +
 \nonumber \\
 & &
2 ({\bf 1},\overline{\bf 15},{\bf 1};{\bf 1},{\bf 1},{\bf 1};\gamma^{1/2})
+
\nonumber \\ 
& & ({\bf 35},{\bf 1},{\bf 1}; {\bf 1},{\bf 1},{\bf 1};\gamma^{-1/2})
+({\bf 1},{\bf 35},{\bf 1}; {\bf 1},{\bf 1},{\bf 1};\gamma^{-1/2})
+ 
({\bf 1},{\bf 1},{\bf 15}; {\bf 1},{\bf 1},{\bf 1};\gamma^{-1/2})
\nonumber \\
55 & : & 2({\bf 1},{\bf 1},{\bf 1};{\bf 15},{\bf 1},{\bf 1};\gamma^{1/2})
+({\bf 1},{\bf 1},{\bf 1};{\bf 6}, \overline{\bf 6},{\bf 1};\gamma^{-1/2})
+ ({\bf 1},{\bf 1},{\bf 1};\overline{\bf 6},{\bf 1},\overline{\bf 4};
\gamma^{-1/2}) +
\nonumber \\
& &
2({\bf 1},{\bf 1},{\bf 1};\overline{\bf 6},{\bf 1},{\bf 4};\gamma^{1/2})
+
 ({\bf 1},{\bf 1},{\bf 1};{\bf 1},{\bf 6},{\bf 4};\gamma^{-1/2})
+2({\bf 1},{\bf 1},{\bf 1};{\bf 1}, {\bf 6},\overline{\bf 4};\gamma^{1/2})
+ \nonumber \\
& & 2({\bf 1},{\bf 1},{\bf 1};{\bf 1},\overline{\bf 15},{\bf 1};\gamma^{1/2})
+ \nonumber \\
& & ({\bf 1},{\bf 1},{\bf 1};{\bf 35},{\bf 1},{\bf 1};\gamma^{-1/2}) 
+({\bf 1},{\bf 1},{\bf 1};{\bf 1},{\bf 35},{\bf 1};\gamma^{-1/2}) 
+({\bf 1},{\bf 1},{\bf 1};{\bf 1},{\bf 1},{\bf 15};\gamma^{-1/2})
\nonumber \\
95 +59 & : &
({\bf 6},{\bf 1}, {\bf 1};{\bf 6}, {\bf 1}, {\bf 1};\gamma^{1/2})
+(\overline{\bf 6}, {\bf 1}, {\bf 1};{\bf 1},{\bf 1}, {\bf 4};\gamma^{1/2})
+({\bf 1}, \overline{\bf 6}, {\bf 1};{\bf 1}, \overline{\bf 6},{\bf 1}
;\gamma^{1/2})+
\nonumber \\
& & ({\bf 1}, {\bf 6}, {\bf 1};{\bf 1},{\bf 1}, \overline{\bf 4};\gamma^{1/2})
 +
({\bf 1}, {\bf 1}, {\bf 4};\overline{\bf 6}, {\bf 1}, {\bf 1};\gamma^{1/2})
+({\bf 1}, {\bf 1},\overline{\bf 4};{\bf 1},{\bf 6}, {\bf 1};\gamma^{1/2})
\label{dbranemodel3}
\ea
The $U(1)$ charges are suppressed but easily obtained.
One finds that the mixed $U(1)$ anomalies
are not universal.

The
$Z_6$ discrete anomalies can be computed
as in the previous models. The charges of
the states under a $Z_6$ rotation of the third
torus are indicated in (\ref{dbranemodel3}).
Here the anomalies
are universal. In particular,
$(Z_6 SU(6)^2_i, Z_6 SU(4)^2_j)=\gamma^2 \times (1,1) $.
These may be canceled by assigning a shift to both the
dilaton axion which couples to the D9 brane gauge bosons
and to the axion of the
$T_3$ modulus which couples to the D5 brane
gauge bosons.

As in the previous examples,
one can also consider the non--$R$ $Z_6$ symmetry.
One finds that this symmetry has a non-universal
anomaly
\beq
(Z_6 SU(6)^2_i,Z_6 SU(4)^2_j) =(\delta^{3},1)~.
\eeq
Again, one may redefine this discrete symmetry to 
include a $U(1)$ factor such that the above anomaly is 
universal. 

\section{CPT} 
\label{cptsection}

In this section we study the $C$, $P$ and $T$ 
symmetries of M--theory in a number of 
backgrounds. 
We first 
consider the low--energy limits
given by a supergravity theory in 
various backgrounds. Because these
theories are local and polynomial in fields,
they respect a CPT symmetry\cite{zumino},
which we identify, along with
other discrete symmetries.

We wish to ask whether these symmetries of the low--energy 
theory are indeed exact. For backgrounds which
are believed to be described non-perturbatively by matrix models,
this is a straightforward exercise. 
The theories we examine in detail are 
M--theory on flat eleven--dimensional space 
(section \ref{mtheory}), M--theory on $T^3$ (section \ref{m3torus}), 
and M-theory on $S_1/Z_2$ (section \ref{ms1z2}), along 
with their corresponding matrix theory descriptions. 
In section \ref{m5torus} we comment on  
M--theory on $T^5$. We only 
discuss matrix models that describe the supergravity and membrane 
dynamics. Whether $CPT$ still exists when the M5 branes 
are included is beyond the scope of this paper, 
and is left for future work. 

In all the examples that we examine, 
we find that the discrete 
symmetries of the classical supergravity theory 
can be found in 
the corresponding matrix models. This includes 
$CPT$. Typically it is either a T or PT symmetry 
of the matrix model that corresponds to 
the $CPT$ symmetry of the space-time 
theory. 

We also emphasize 
that we focus on theories that are 
Lorentz invariant. 
A violation of 
CPT in a Lorentz invariant theory 
is much more non--trivial, and interesting.  

\subsection{Chern--Simons Theory in Five Dimensions}

It
is instructive to start by
discussing the discrete symmetries
of a five--dimensional Chern-Simons theory coupled
to a current. This theory has some similarities 
to the eleven--dimensional theory considered 
in the next section, but since it is simpler 
we begin here first.  

Electromagnetism in five
dimensions coupled to a current preserves 
three symmetries: $C$, $T$, and $P$.
Since here there are an even 
number of spatial dimensions, 
the parity symmetry
corresponds to a reflection about an odd
number of spatial
directions.
The addition of a Chern-Simons coupling
violates both the charge conjugation and parity symmetries. 
But $T$ and the combination $C P$ are preserved. 

\subsection{M--Theory in Eleven--Dimensions}
\label{mtheory}

\subsubsection{Discrete Symmetries of the Classical Theory}

The low-energy limit of a supersymmetric quantum
theory in
eleven--dimensions is described by eleven--dimensional
supergravity. It contains a graviton, gravitino and 
also a
three-form potential $C_{3}$ that has a Chern--Simons coupling.
The Lagrangian is, schematically \cite{juliacremmer}, 
\ba
{\cal L} & = & - eR - eG^2 +  ie \overline{\psi}_{\mu} \Gamma^{\mu \nu \rho}
D_{\nu} \psi_{\rho}
\nonumber \\
& & +C \wedge G \wedge G
- ie\left(\overline{\psi}_{\lambda} 
\Gamma^{\mu \nu \rho \sigma \lambda \tau} \psi_{\tau} 
+ \overline{\psi}^{\mu} \Gamma^{\nu \rho} \psi^{ \sigma}
\right)(G + \hat{G})_{\mu \nu \rho \sigma}  ~.
\ea
Here $G=dC_{3}$ is the four-form field strength, 
$\psi_{\mu}$ is 
the gravitino, 
$D_{\mu}$ is the covariant derivative 
including the spin connection and $e$ is the determinant of
the vielbein. 
$\hat{G}$ 
is a combination of the four--form field strength 
and a term involving two gravitinos. Under 
$C$,$P$ and $T$ 
it transforms like $G$. 
The $\Gamma$ matrices will be chosen to 
be real, with $\Gamma^0$ antisymmetric and all the others 
symmetric. $\Gamma$'s with $n$ Lorentz indices are 
given 
by the antisymmetric product of $n$ $\Gamma$ matrices and 
are all real.  
Finally, $\overline{\psi} \equiv \psi^{\dagger} 
\Gamma^0$. 

From the previous example we expect the eleven-dimensional
theory to preserve only $T$ and $CP$.
Under $T$ the three-form potential and field-strength 
transform 
as  pseudotensors.  
Under $CP$ the three-form potential 
transforms as a tensor under the reflection, 
and acquires an overall $(-)$ sign from the 
charge conjugation. 
It is straightforward to
confirm that the
bosonic part of the action is invariant under either of
these symmetries. 

Verifying that the supersymmetric theory is 
$CP$ and $T$ invariant requires a bit more work. Under $CP$ 
(a reflection in $x^i$ say), a Majorana $SO(10,1)$ spinor 
transforms 
\beq 
\theta(x,x^i) \rightarrow \Gamma^i \theta(x,-x^i)~.
\eeq
Then 
\beq 
\overline{\theta} \Gamma_{\mu_1 \cdots \mu_n} \theta 
\eeq 
transforms as a tensor when $n$ is odd and as 
a pseudo-tensor when $n$ is even. 
This pseudotensor property is crucial to make the action 
$CP$ invariant. 

Under $T$, a spinor transforms as 
\beq 
\theta(x^0,x^i) \rightarrow \Gamma^0 \theta (-x^0,x^i), 
\eeq 
with the gravitino transforming as a vector-spinor. One may check 
that all the interactions are $T$ invariant. 
This is because 
under $T$ 
\beq 
i \overline{\theta} \Gamma^{\mu_1 \cdots \mu_n} \theta
\eeq
transforms as a tensor when $n$ is odd and as 
a pseudo-tensor when $n$ is even.  

The classical theory also has membrane solutions. 
The Lagrangian for the membrane in flat space is \cite{dewitt1}
(dropping numerical factors)
\beq
{\cal L} = \sqrt{g(x,\theta)} + i \epsilon^{ijk}
\left( \partial_i X^{\mu}(\partial _j X^{\nu} + i \overline{\theta} 
\Gamma^{\nu} \partial_j \theta) + 
\overline{\theta} \Gamma^{\mu} \theta 
\overline{\theta} \Gamma^{\nu} \theta \right) \overline{\theta} 
\Gamma_{\mu \nu} \partial _k \theta.
\eeq 
Here $X^{\mu}(\tau,\sigma^{i})$ parameterize the position 
of the membrane, and $\theta(\tau,\sigma^{i})$ is a 
Majorana spinor which has 32 real components. 

This action has both a $CP$ and a $T$ invariance if we extend 
the symmetries to include a world-sheet parity and world-sheet 
time-reversal.

First consider $CP$. Since $\overline{\psi} \Gamma_{(2)} \psi$ transforms 
as a pseudo-tensor whereas $\overline{\psi} \Gamma_{(1)} \psi$ 
transforms as a vector, none of the fermion terms 
are invariant. 
Also note that 
the fermion terms have an odd number of world--sheet 
derivatives, so  
the Lagrangian is not invariant under a world-sheet parity 
transformation either. These two transformations 
can be combined so that the action is invariant 
under the spacetime $CP$ and a world-sheet 
parity transformation. 

Next consider $T$. The membrane Lagrangian is invariant 
under $T$ if the world-sheet time $\tau$ is also flipped. 
This too is reasonable. For the above Lagrangian, although covariant, 
has a large gauge symmetry which includes (proper) reparameterization 
invariance. This redundancy is fixed in the light-cone 
frame, where $X^{+} = \tau$. A time-reversal in $X^0$ implies 
a time-reversal in the world-sheet time.

Finally the coupling of the membrane to
the three-form potential is given by
\beq
\int _{M_2} C_3 =\int_{M_2} dt d^2 \sigma
C _{\mu \nu \gamma} {d x^{\mu} \over d t }
\left\{ X^{\nu}, X^{\gamma} \right \}_{PB} ~.
\eeq
Using the information provided above
one finds that these too are invariant under $T$ and $CP$. For 
under $CP$ or $T$ the three--form potential transforms as a 
pseudo-tensor. For $CP$ we must also include a 
world-sheet parity reflection which causes 
the Poisson--bracket to transform as pseudo-tensor. 
The current--coupling is then invariant. For 
$T$ we must also flip 
the world-sheet time. 
The velocity $\dot{X}^{\mu}$ transforms 
as a pseudo-vector (that is, as a momentum), 
whereas the Poisson bracket 
now transforms as a tensor. The membrane current--coupling 
is then $T$ invariant. 

In summary, the eleven-dimensional supergravity theory,
including the membrane dynamics, is invariant under
two discrete symmetries: $T$ and $CP$. 
We leave the issue of whether 
these symmetries are preserved when M5 branes are 
included to future research, but we expect that the answer
is yes.
It is natural to then ask whether these symmetries
are preserved in the exact quantum theory. 
This leads us
to matrix theory, which 
is conjectured to be an exact quantum
gravity theory whose low-energy limit is
eleven-dimensional supergravity, including the membrane 
excitations. 

\subsubsection{Matrix Model}

The matrix model is given by the truncation of
$N=1$ $d=10$
Super-Yang-Mills $U(N)$ theory down to zero dimensions.
The large $N$
limit of
this quantum mechanical system is conjectured \cite{matrixtheory} to
describe M-theory in eleven--dimensional Minkowski space,
while for finite $N$ it is believed to give the discrete light cone
quantization of the theory.

We shall see that 
the matrix model has two discrete symmetries,
C$_{\hbox{M}}$ and T$_{\hbox{M}}$. These correspond to the
charge and parity-time reversal symmetries of the ten-dimensional
minimal super-Yang Mills theory. As we shall see,
the correspondence 
we will establish is:
\ba
\hbox{11 dimensions} & & \hbox{Matrix theory} \\ \nonumber
 \hbox{\em C R}_{10} \hbox{\em T} 
~~~~~~ & \longleftrightarrow &  ~~~~~~\hbox{T}_{\hbox{M}}  \\ \nonumber
\hbox{\em C R}_1 \cdots \hbox{\em R}_9 
~~~~ & \longleftrightarrow &  ~~~~~~\hbox{C}_{\hbox{M}}  
\ea
where $R_i$ reverses the sign of the $i$'th coordinate.
{\em Note that CPT in the eleven-dimensional theory 
is equivalent to a time-reversal symmetry of the 
matrix model.}
To minimize confusion, discrete symmetries of the eleven-dimensional 
or space-time 
theory will be italicized, while those of the matrix model 
will be given in Roman type. 

The Lagrangian is
\beq
L={1 \over 2} \hbox{tr} \left(
D_t {\bf X} D_t {\bf X}  + [X^k,X^l]^2 +
i {\theta} D_t {\theta} -\theta \gamma^k [X_k,\theta] \right) ~.
\label{matrixmodelL}
\eeq
Here $X^k$ and $\theta$ are $N\times N$ hermitian matrices, with
$k=1 \ldots 9$.
$\theta$ is an anti-commuting $SO(9)$
spinor with 16 real components.
With
the hermitian conjugation
of two anti-commuting variables defined to be $(\lambda \eta)
^{\dagger}
= \eta^* \lambda^*$, the two fermionic terms in the Lagrangian
are hermitian.
The $\gamma^i$ matrices are $16 \times 16$ matrices
satisfying the
$SO(9)$ Clifford algebra $\{\gamma_i,\gamma_j\}=2 \delta_{ij}$
and are all real and symmetric.

We recognize the first two terms as the ten-dimensional 
gauge field strength, and the last two as the 
gaugino kinetic term and current coupling.

This theory has two real sixteen component supercharges,
which are \cite{bankslectures}:
\ba
q_{\alpha} &=& \hbox{tr}  \theta_{\alpha}
\nonumber \\
Q_{\beta} &=& \hbox{tr} \left( P^i \gamma^i _{\beta \alpha} \theta_{\alpha}
+ i\gamma^{ij}_{\beta \alpha} [X^{i}, X^{j}] \theta_{\alpha} \right) ~.
\label{qs}
\ea
Here $\gamma^{ij} \sim  [\gamma^i,\gamma^j]$ is purely real.
For future reference,
\beq
\{q_{\alpha},Q_{\beta} \} \sim \gamma^i _{\alpha \beta}
\hbox{tr} P_i + Z_{ij} \gamma^{ij} _{\alpha \beta} ~,
\label{anticom}
\eeq
where 
\beq
Z_{kl} = i \hbox{tr}[X^k, X^l] ~.
\label{m2charge}
\eeq
is the charge of a membrane stretched in the $i$-th and 
$j$-th directions \cite{banksmatrices}. 

We next discuss the C$_{\hbox{M}}$ and T$_{\hbox{M}}$ 
symmetries of this quantum 
mechanical model. 

{\bf C}$_{\hbox{M}}$: As 
asserted above, the matrix model has a charge conjugation
symmetry. In fact there are nine \cite{motl}.
These descend from the charge conjugation symmetry
existing in ten-dimensions and the $SO(9)$ rotation group.
Although the ten-dimensional theory is chiral,
a charge conjugation symmetry can still be imposed.
This is possible in $d=4k+2$ dimensions,
but not in $d=4k$ (which includes the familiar four dimensions)
\cite{polchinskibook}. A Majorana condition can also be
imposed if the representation is real. In this case it is,
since the fermions are in the adjoint representation.
For our purposes it is sufficient to 
study the charge conjugation symmetry which 
is:
\beq
{X}_i (t) \rightarrow
-{X}^{T}_i (t) ~, 
\theta(t) \rightarrow  \theta^{T}(t)
\label{Ci}
\eeq
with the transpose acting on the $U(N)$ indices. Normally
the spinors are conjugated, but this is trivial
since the spinors are real.

Since commuting diagonal matrices $X^i_{aa}$
describe the transverse
center-of-mass coordinate of a particle in the light-cone, this
charge conjugation describes
the reflection (not connected to the identity)
$R_1 \cdots R_9$ in eleven-dimensions. 
The momenta also transform correctly,
appropriate for a reflection.
This transformation also induces $C$ in the space-time theory. 
The matrix model expression for 
the charge of an infinite M2 brane at rest 
is given by 
(\ref{m2charge}). 
Since under $C_{\hbox{M}}$ the matrices are transposed, 
the charge of the membrane is flipped. 
This is also confirmed by inspecting 
the supersymmetry algebra (\ref{qs}) and 
(\ref{anticom}). 
For under $C_{\hbox{M}}$,   
$Q \rightarrow -Q$, and consistency of the 
algebra requires 
that $Z \rightarrow -Z$. 

{\bf T}$_{\hbox{M}}$: This is just the PT transformation
of a zero mode gauge field in ten--dimensions.
In general, 
this is described by 
antiunitary operator $K$ that  
sends $t $ to $-t$ and 
has also acts 
on the 
variables of the model - in this case, $N \times 
N $ matrices.
For a non-relativistic electron, for example, 
$K = \sigma_y K_0$ where $K_0$ complex conjugates 
the electron wavefunction.  
Here we {\em define} time 
reversal by the antiunitary operator 
$K$ that 
acts on the $N\times N$ matrices by: 
\beq
K X_{i}(t) K^{-1}
= X_{i}(-t) ~~,  ~~ K \theta(t) K^{-1} = \theta(-t) ~~,
K c K^{-1} = c^* ~.
\eeq
Here $c$ is any complex number that is {\em not} in 
the $N \times N$ matrices $X$ or $\theta$. 
We later demonstrate that K defined in this 
way acts on the supergraviton states
in a way that agrees with the
$C P_{10} T$ transformation 
in the eleven-dimensional theory. To begin 
though, it   
is straightforward to verify that the 
matrix model lagrangian is invariant under 
this transformation, in that $K L(t) K^{-1} = L(-t)$. 

Next we describe the eleven-dimensional 
interpretation of this symmetry. 
First note that the transverse coordinates
transform rather trivially. 
The transverse
momenta, however, do change sign. Both of these 
are what we
expect of an eleven-dimensional time-reversal.
{\em But the
eleven-dimensional interpretation
requires an inversion
of both of
the light-cone coordinates}.
This is because the quantum mechanical mass,
which is the longitudinal momentum, is invariant.
This means that the matrix model T$_{\hbox{M}}$
symmetry reverses both light-cone directions, 
corresponding at least 
to a $P_{10} T$ symmetry in eleven dimensions.
Finally, this symmetry also implies a 
charge conjugation in eleven-dimensions. 
For under K the charge (\ref{m2charge})
of a membrane transforms as $Z_{ij}\rightarrow -Z_{ij} $.
To keep the current coupling invariant 
this requires a charge conjugation of the three-form.

We can
also inspect the
expression for the supercharges to find that under T$_{\hbox{M}}$
the Hamiltonian is invariant. 
This is sufficient to establish that T$_{\hbox{M}}$ is 
a symmetry of the quantum mechanical model. 
Further, since it is anti-unitary, 
it will exchange ingoing and outgoing states. 
From the preceding discussions, 
T$_{\hbox{M}}$ corresponds to the
$C R_{10} T$ symmetry of the eleven--dimensional 
theory, which since eleven-dimensions
has an even number of spatial directions, {\em is}
the $CPT$ symmetry. Assuming eleven-dimensional
(proper) Lorentz invariance, this symmetry is unique.

So far we have demonstrated that the matrix model 
T$_{\hbox{M}}$ symmetry has the correct interpretation as a $CPT$ 
symmetry
in eleven dimensions, and is a symmetry of the matrix 
model 
hamiltonian. 
It remains to establish that 
it acts appropriately on the states of the theory.

We have already seen that K reverses the transverse 
momenta, preserves the longitudinal 
momentum,  and also charge conjugates. But a $C P_{10} T$ 
transformation in eleven dimensions should also 
change the helicity of the supergraviton states. 

To see this, we focus on states that have zero transverse momenta. 
(But in eleven dimensions they have non-zero longitudinal 
momentum.) The helicity of these states is then
given by their charge under $U(1)$ subgroups of the $SO(9)$ little group.
 
In addition, we study the asymptotic properties 
of the supergraviton states in an $N=2$, $U(2)$ model. 
Although the Hilbert space for this model has only two 
supergravitons, it is large enough for our purposes. 
Plefka and Waldron \cite{plefkawaldron} have discussed the 
construction and computation of scattering 
amplitudes in this model. We follow 
their construction and notation. 

For $N=2$ we expand 
\beq 
X_i = X^0 _i {\bf 1} + \vec{X} _i \cdot \vec{\bf T} 
\eeq
\beq 
\theta_{\alpha} = \theta^0 _{\alpha} {\bf 1}
+ \vec{\theta} _{\alpha} \cdot \vec{\bf T} 
\eeq
where $X^0_i$ describes the center-of-mass position of 
the supergraviton with fermionic zero modes $\theta^0$, 
and $\vec{X}_i$ and its superpartner describe the 
relative separation. Vectored quantities denote 
values in the $SU(2)$ algebra. 
The $U(1)$ part 
corresponds to the center-of-mass motion, 
and variables in the $SU(2)$ part describe 
the relative separation of two supergravitons.

In terms of these variables the Hamiltonian has a 
simple form, 
\beq 
H= P^0_i P^0_i + \vec{P}_i \cdot \vec{P}_i  
+(\vec{X}_i \times \vec{X}_j)^2 + i \vec{X}_j \cdot 
(\vec{\theta} \times \gamma_j  \vec{\theta}) \ ~,
\eeq
and the action of K is given by 
\beq 
KZ^IK^{-1}=Z^I ~~, I=0,1,3 ~~,~~ K Z^I K^{-1} =- Z^I ~~,I=2, 
~~KcK^{-1} = c^* ~,
\eeq 
where 
the label $I=0,1,2,3$ refers to the generators of $SU(2)$, 
and $Z= X$ or $\theta$. 
It can be verified that the Hamiltonian is invariant.

To discuss the construction of the states, 
it is more convenient to only 
keep invariance under 
the $SO(7) \times U(1)$ subgroup of the $SO(9)$ little 
group manifest, as described in 
\cite{dewitt1,plefkawaldron}. The $U(1)$ generator $J_{89}$ 
gives 
the helicity of a state in the 89 plane. 
Under $SO(7) \times U(1)$, the 16 dimensional spinor $\theta^0$ decomposes 
into two 8 dimensional $SO(7)$ spinors $\theta^0 _+$ 
and $\theta^0 _-$ that differ by their $SO(8)$ chirality. 
They can be organized into $\lambda= 
\theta^0_+ + i \theta^0 _-$ and its conjugate $\lambda^{\dagger}$. 
These obey the algebra   
$\{\lambda_{\alpha}, \lambda^{\dagger}_{\beta}\} 
=\delta_{\alpha \beta}$. The supergraviton multiplet 
is constructed by applying the raising operators 
$\lambda^{\dagger}$ to the ground state $|- \rangle$ 
which is annihilated by all the lowering operators. 
This state has $U(1)$ charge $-1$. 
A transverse 
graviton state is, for example \cite{plefkawaldron}, 
\beq 
 e^{i k_i X^0 _i} 
h_{ij} (\lambda^{\dagger} \gamma_i \lambda^{\dagger} 
)(\lambda^{\dagger} \gamma_j \lambda^{\dagger})|- \rangle
\label{supergravstate}
\eeq 
with the $\gamma_i$ matrices real antisymmetric 
$SO(7)$ Dirac matrices.
A transverse 3-form state is \cite{plefkawaldron}
\beq 
 e^{i k_i X^0 _i} 
c_{ijk} (\lambda^{\dagger} \gamma_{[ij} \lambda^{\dagger} 
)(\lambda^{\dagger} \gamma_{k]} \lambda^{\dagger})
|- \rangle
\eeq 
and so on. As an aside, we note 
that given this explicit expression for  
the three--form state, it can be verified that 
under $C_{YM}$ this state 
transforms as $c_{ijk} \rightarrow c_{ijk}$ with the  
transverse momentum reversed. This is indeed the 
correct $C R_{1} \cdots R_{9}$ transformation of 
the three--form. 

To discuss the action of
T$_{\hbox{M}}$ it is convenient to expand our state 
into modes with definite helicity. 
The basis appearing in (\ref{supergravstate}) 
is not well suited for this purpose, 
since it corresponds to a real basis. 
But it is straightforward to construct the helicity 
basis corresponding to states with definite 
charge. 
For instance, focus on an $SO(3)$ subgroup 
of $SO(7)$. From a vector of $SO(3)$, we can 
construct states with charges $m_{12}=0,\pm1 $ under
$J_{12}$. In the matrix model this corresponds 
to the states 
\ba
|B_{\pm} \rangle 
&=& \lambda^{\dagger}(\gamma_1 \pm i \gamma_2 ) \lambda^{\dagger} 
|- \rangle \nonumber \\ 
|\tilde{B}_{\pm} \rangle &=& \lambda(\gamma_1 \pm i \gamma_2 ) \lambda 
|+ \rangle 
\nonumber \\
|B_0 \rangle &=& \lambda^{\dagger}\gamma_3 \lambda^{\dagger} 
|- \rangle \nonumber \\
|\tilde{B}_0 \rangle &=& \lambda \gamma_3 \lambda 
|+ \rangle 
\ea
with $\lambda^{\dagger} |+ \rangle=0$. In addition, 
the states with $|- \rangle$ $(|+ \rangle)$ have charge 
$m_{89}=+1$ $(-1)$ under $J_{89}$. 
Since $K$ acts as 
\beq 
K|- \rangle = |+ \rangle ~~, ~~K \lambda K^{-1} = \lambda^{\dagger}
\eeq 
we see that 
\beq 
K|B_{\pm} \rangle = |\tilde{B}_{\mp}\rangle ~~, K |B_0 
\rangle = |\tilde{B}_0 \rangle ~. 
\eeq
That is, $K$ changes a state $(m_{12}, m_{89})$ 
into another having the opposite helicities. 
It is clear that 
we can use this method to construct, for example,  
a ${\bf 5}$ which will have well--defined 
helicity $m_{12}=-2, \ldots 2$. These describe 
a graviton with polarizations in the transverse 
$1,2,3$ directions. It is also clear 
that as before, $K$ will change a state with helicity 
$m_{12}$ into one with $-m_{12}$. Thus $K$ 
acts on the transverse supergraviton 
states in a way that is appropriate for a $CPT$ transformation 
in eleven--dimensions.

This section concludes with some remarks summarizing
what has been established. Matrix theory has two
discrete symmetries, which in the eleven--dimensional
interpretation correspond to the discrete
symmetries of the low-energy theory,
$CPT$ and $CP$. It is curious that the $CPT$ symmetry of
the eleven-dimensional theory does not correspond
to the CPT  symmetry of matrix model, inherited
from ten dimensions. Rather, it is PT. We do not
understand why it turned out this way.

But
which dynamical processes, viewed from
eleven-dimensions, will preserve
these symmetries?
It is conjectured that matrix theory is complete in
the following sense \cite{matrixtheory}. It should describe
any configuration or dynamics that can be
produced by finite energy
multigraviton scattering.
This
includes supergraviton scattering at low-energies, but
also more exotic (and interesting) processes.
For instance, since membranes or fivebranes of finite
size (and energy) carry no charge, they can be produced
by multigraviton scattering. These are not static
configurations, since they eventually collapse to form
non-extremal black holes. 
The process leading from
(super)gravitons in the initial state to the
(super)gravitons produced by
the decaying black holes
will preserve $CPT$ and $CP$.

But there are other notable degrees of freedom - membranes
and
fivebranes of infinite size. These cannot be
produced by supergraviton scattering, so a
separate argument is needed to argue that
their dynamics preserves $CPT$ and $CP$.
Matrix theory descriptions
exist for the infinite membrane and antimembrane\cite{matrixtheory},
so processes in these 
backgrounds will preserve $CPT$ and $CP$.

When we come to
fivebranes of infinite size our arguments are incomplete.
This is because matrix theory does not
describe all fivebranes.
There do exist matrix configurations
for fivebranes carrying a
fivebrane and membrane charge \cite{banksmatrices}, and these will
obey $CPT$ and $CP$.  But
there are other fivebranes that
require a
different matrix theory. For instance, 
$k$ longitudinal fivebranes
are described \cite{berkoozdouglas}
by the dimensional truncation to zero dimensions
of $N=2$, $d=4$
super-Yang-Mills with $k$ fundamental
hypermultiplets.

We also lack arguments for 
transverse fivebranes (of infinite size).
By Seiberg's
argument \cite{seibergmatrix}
they are
described by the six dimensional $(2,0)$
little string theory. This is not a local
quantum field theory, so there is no argument for
the existence of a $CPT$ symmetry.
It would be 
interesting to investigate this question 
further in the recent matrix model proposal for 
transverse five--branes \cite{fivebranerecent}.

\subsection{M-Theory on T$^3$}
\label{m3torus} 

\subsubsection{Discrete Symmetries of the Classical Theory}

At the classical level the eight--dimensional 
model we consider is given by the compactification of 
11-dimensional supergravity and membranes  
on a three-torus. (As in our discussion 
of the eleven--dimensional theory, here we  
focus on backgrounds that do not include M5 branes.) 
The discrete symmetries we focus on are those 
inherited from the eleven--dimensional theory.
Recall that the eleven-dimensional
theory has two discrete symmetries: $T$ and
$CP$. For $CP$, we may choose the reflection to occur in one 
of 
the seven non--compact directions. For instance, this 
could be the longitudinal direction ($R_{10}$) of the light-cone 
gauge. 
But for suitable choice of torus there are also three more parity symmetries
$P_i$,
which combine a reflection in one of the internal
directions ($R_7$, $R_8$ or
$R_9$, say),  with a reflection in all of
the non-compact directions.
They correspond to unbroken discrete elements of the
$SO(10)$ rotation group in eleven--dimensions. 
As the product of any two 
of these is a discrete rotation in the internal three-torus, 
the three parity symmetries, together with the $CP$ 
symmetry, can be combined to give a $C$ and $P$ symmetry, 
and two internal symmetries. 
In total, the 7+1 dimensional theory has
five independent discrete symmetries :
$P$, $C$, $T$ and two internal discrete symmetries.
The product of $CP$ and $T$ is the
$CPT_8$ symmetry of the eight--dimensional theory.

These symmetries may be spontaneously broken 
by the vev of a field. An interesting 
example is if $C_{789}$ 
is non--zero. This breaks the eleven--dimensional 
$CPT$, but still preserves the eight--dimensional 
$CPT$ and Lorentz invariance.

In the next section we examine  
the matrix model for M theory in this 
background, with and without 
$C_{789}=0$, and 
ask if there is an unbroken symmetry which 
corresponds to the eight--dimensional $CPT$. 
In both cases the answer is yes.

\subsubsection{Matrix Theory Description}

By Seiberg's argument \cite{seibergmatrix},
the matrix theory description of M-theory on $T^3$ in
the light-cone is
given by the large $N$ limit
of the $d=3+1$, ${\cal N}=4$ supersymmetric
$U(N)$ theory compactified on the dual three-torus
$\widetilde{T}^3$.
As we now show,
the discrete
symmetries of this theory
correspond to those found in the
classical 7+1 dimensional
theory.

The couplings of the 
four--dimensional theory are the gauge coupling 
and the theta parameter. We begin by first setting  
the theta parameter to zero. 

Then
the gauge theory has separate
CP$_{\hbox{YM}}$, C$_{\hbox{YM}}$ and T$_{\hbox{YM}}$ symmetries.
As in the original 
matrix model example, the interpretation 
of these symmetries in the eight--dimensional 
theory is non--trivial. 

Thinking
about this theory as arising from
the dimensional truncation of a
ten--dimensional Yang-Mills theory, one
might have expected only two discrete symmetries.
The four--dimensional C$_{\hbox{YM}}$ and PT$_{\hbox{YM}}$
are inherited directly
from the ten-dimensional C$_{\hbox{YM}}$ and PT$_{\hbox{YM}}$ symmetries.
But there is an additional
discrete symmetry in the four--dimensional theory, which was 
part of
the Lorentz group in ten dimensions.
Four--dimensional P$_{\hbox{YM}}$
is given by a reflection of the 
three non--compact directions together with one of  
the internal
directions.

On
the $\tilde{T}^3$, the rotation group 
is broken to the discrete subgroup of rotations preserved 
by the three-torus. We may combine P$_{\hbox{YM}}$ with these 
discrete symmetries to obtain 
three
parity symmetries P$^i$$_{\hbox{YM}}$, each corresponding to a
reflection in one of the directions of the dual
three-torus.
The Yang--Mills theory also has a
$SU(4)$ symmetry, which in the eight--dimensional 
theory corresponds to the manifest rotation symmetry
in light-cone gauge.

To obtain the mapping between the symmetries 
of the Yang-Mills theory and the supergravity 
theory, we need to discuss the BPS states of 
the two theories \cite{mt3}.  In the Yang-Mills 
theory we can have field configurations 
with an electric field $\vec{E}$, magnetic field $\vec{B}$, 
and/or a non-zero vev for the six scalar adjoints $X^i$. 
On the moduli space the six scalars parameterize 
the position of the  (super)graviton in the (non-compact) transverse 
six dimensions. Field configurations with quantized 
electric flux correspond to the KK 
momentum of supergravity states. Field 
configurations with magnetic flux 
correspond to transverse 
membranes (i.e., not along the light--cone direction 
$R_{10}$) that are 
wrapped around two one--cycles 
of the three-torus. The winding number of the membrane 
corresponds to the quantized magnetic flux.

We discuss each of these in turn. 

{\bf C$_{\hbox{YM}}$}: This flips all of the transverse scalars. Since 
there are six of these, this is just an element of 
$SO(6)$. But since the electric field also 
changes, in the supergravity theory 
this must correspond to flipping 
all the components of the KK momentum. Thus C$_{\hbox{YM}}$ includes 
a reflection in 7,8 and 9. 
But under C$_{\hbox{YM}}$ the magnetic field changes 
as well. In the supergravity theory 
this corresponds to flipping the winding number 
of the membrane. But the winding number is 
just the charge of the membrane. 
So C$_{\hbox{YM}}$ includes a charge 
conjugation in the supergravity theory. 
Therefore 
C$_{\hbox{YM}} \leftrightarrow$ $C R_7R_8R_9$.

To see this more formally  
we have to discuss how the charge of a membrane 
transforms under 
$CP$. 
The charge or winding number of a membrane wrapped 
around cycles of the torus in the $X^i$ 
and $X^j$ 
directions, with $i,j$ one of 7,8,9, is given by 
\beq 
Z_{ij} \sim \int d^2 \sigma \{X^i, X^{j}\}_{PB} ~.
\label{m2longcharge}
\eeq  
Under a reflection in 7,8 and 9 the charge is invariant. 
But under a $CP$ transformation ($CR_{7}R_{8}R_{9}$)
the orientation of the membrane is also changed. 
This $CP$ changes the charge, which agrees with the 
corresponding change in the Yang-Mills magnetic field. 
So we have learned that C$_{\hbox{YM}} 
\leftrightarrow CR_7R_8R_9$. 

{\bf PT$_{\hbox{YM}}$}: This reverses the Yang--Mills time coordinate, 
which is the light--cone time coordinate of the 
M--theory. So PT$_{\hbox{YM}}$ implies $TR_{10}$. Now under 
PT$_{\hbox{YM}}$ both the electric and magnetic fields change. A 
sign change in the electric fields means 
that in the 
M--theory all the KK momenta change. The 
action of $T R_{10}$ alone does this, 
so no additional reflection 
in the internal space is required. 
That the magnetic field changes 
means that in the M--theory 
the membrane charge changes. But 
(\ref{m2longcharge}) implies that the membrane 
charge does not change sign under  
$T R_{10}$ alone. So to  
obtain the correct winding number transformation, 
PT$_{\hbox{YM}}$ must also imply 
a charge conjugation in the
M--theory, which acts to reverse the membrane orientation. 

Note that again it is PT$_{\hbox{YM}}$, and not CPT$_{\hbox{YM}}$, 
that corresponds to the eight--dimensional CPT symmetry. 
This can be better understood if 
we try, given the symmetries of the supergravity 
theory, to see what it could have been. 
Since a time reversal 
is involved, the Yang--Mills parity--time 
reversal symmetry could {\em a priori} only correspond 
to $T$, $CT$, $T R_{10}R_{7}R_{8}R_{9}$, 
$R_{10} T$, 
or $CR_{10} T$ (=$CPT$) of the eight--dimensional theory. 
The arguments of the preceding paragraph exclude 
all but the last possibility.

{\bf P$^i_{\hbox{YM}}$}: In the ten--dimensional theory this 
was an element of the $SO(9)$ rotation group. 
Therefore 
one and only one of the adjoint scalars transforms.
A change in one of the adjoint scalars corresponds in  
the supergravity theory 
to a parity reflection in one of the transverse non--compact 
directions. Therefore P$^i_{\hbox{YM}}$ implies 
$P_i$ of the M--theory. 
Returning to the Yang-Mills theory, 
P$^i_{\hbox{YM}}$  
changes the electric field only in the 
$i-$th direction. By the correspondence, the 
graviton momentum along the $i-$th cycle 
should be modified. 
The most natural choice consistent 
with these two observations is 
that this symmetry of the Yang-Mills 
theory is $P_i$ of the M-theory. 
This 
guess also preserves the correspondence 
between the magnetic flux and membrane charge. 
For under P$^i_{\hbox{YM}}$,
$F_{kl}$ does not change for $k$ and $l$ 
not equal to $i$, whereas $F_{ik}$ does change. 
But this is precisely 
the transformation of the winding number 
under $P_i$ : 
flipping only $R_i$ changes $Z_{ij}$ and leaves the  
other components invariant. From this we find that 
P$^i_{\hbox{YM}} \leftrightarrow P^i$. 

To summarize: 
\ba
\hbox{7+1 dimensions} & & \hbox{Matrix theory} \\ \nonumber
 C R_{10} T ~~~~~~ & \longleftrightarrow &  ~~~~~~\hbox{PT}_{\hbox{YM}}  
\\ \nonumber
C R_7 R_8  R_9
~~~~ & \longleftrightarrow &  ~~~~~~\hbox{C}_{\hbox{YM}}   \\ \nonumber 
P_i & \longleftrightarrow & \hbox{P$^i$}_{\hbox{YM}}  \\ \nonumber
\ea
In particular, the eight dimensional CPT symmetry is
PT$_{\hbox{YM}}$ of the Yang-Mills theory.
So far we have assumed 
that the theta parameter of the Yang-Mills 
theory vanishes. If instead it is non--zero, 
then the time and parity
symmetries of the Yang-Mills theory are broken, but
the combinations  
PT$_{\hbox{YM}}$ and C$_{\hbox{YM}}$ are preserved. In addition, 
each of  the P$^i_{\hbox{YM}}$ 
symmetries are broken, but the product of 
any two is not. In total, the theta parameter 
violates only P$_{\hbox{YM}}$ and $T_{\hbox{YM}}$ but preserves 
four discrete symmetries. In particular, 
PT$_{\hbox{YM}}$ is preserved. This 
implies that M--theory in this background 
should 
have an unbroken eight--dimensional $CPT$. 

A non--zero theta term corresponds in the
supergravity 
theory to $C_{789} \neq 0$ 
\cite{mt3}. This is satisfying, 
for this preserves the eight--dimensional 
$CPT$. In fact, one finds that the symmetries preserved 
in the Yang--Mills theory with a non-zero 
theta parameter are indeed, by the correspondence 
given above, the same symmetries 
preserved by $C_{789}$.

To summarize: we have found that the matrix model 
has an unbroken PT symmetry,  and that it corresponds 
to the $CPT$ symmetry of the eight--dimensional 
M--theory. Just as before, we have not studied 
backgrounds including M5 branes. 

\subsection{M-Theory on $S_1/Z_2$}
\label{ms1z2}

In this section, we consider M-theory on $S_1/Z_2$.
The large radius limit of this theory 
is the Horava-Witten model, 
and the small radius limit is the $E_8 \times E_8$
weakly coupled heterotic string. We begin 
by recalling in section \ref{weakheterotic} 
the CPT symmetry of the 
weakly coupled heterotic theory, and 
then in section \ref{hetmatrixsec} 
identify this symmetry in the matrix model 
description. 

\subsubsection{Discrete Symmetries of the Heterotic String}
\label{weakheterotic} 

The heterotic string
has one discrete symmetry $\theta$ which is
$CPT$ in ten dimensions \cite{polchinskibook}.
In the covariant formulation 
$\theta$ acts as $X^{0,1}(\sigma,\tau)
\rightarrow -X^{0,1}(\sigma,\tau)$, where 
$X^{0,1}$ are the two light-cone directions.
In the (RNS) formulation where the world-sheet fermions
$\psi^{\nu}$ 
transform as space-time vectors, $\theta$ acts to 
also change the signs of   
$\psi^{0,1}$. 
All other fields transform trivially,
including the 32 left-moving $\lambda$'s.

In light cone gauge, the
Weyl and diffeomorphism 
invariance of the world-sheet 
may be used to completely fixed these coordinates and 
their fermionic partners to be   
$\psi^-=0$, $\partial_+ X^+ = p^+ /2$ and  
\cite{GSW} 
\ba
\partial _+ X^-&=&(\partial_+ X^i \partial _+ X^i 
+ {i \over 2} \psi^i \partial _+ \psi^i)/p^+ ~,\nonumber \\
\psi^- &=&  2 \psi^i \partial_+ X^i /p^+ ~,
\label{gaugefix}
\ea
along with setting the world-sheet metric to be 
canonical. 
The only residual reparameterization transformations
are constant shifts of the origin of the world-sheet spatial 
coordinate, which lead to the level matching condition.
In  
the light-cone gauge $\theta$ acts as 
$(\sigma,\tau)\rightarrow (-\sigma,-\tau)$ and 
leaves the transverse bosonic and fermionic 
variables invariant: 
\beq 
X^{i}(x^+) \rightarrow X^{i} 
(-x^+)~~,~~ \psi^i(x^+) \rightarrow \psi^i(-x^+)~. 
\eeq
Because of the gauge fixing, this is not a world sheet
diffeomorphism.
In light-cone gauge $\tau \rightarrow \tau^{\prime}(\tau)
=-\tau$ is a combination of 
time-parity reversal in the target space with a reflection 
of the world-sheet coordinates. 
This means that $X_+$ should transform as 
\beq 
X_{+}(\tau,\sigma) \rightarrow X_{+}^{\prime}(-\tau,-\sigma) 
= - X_{+}(\tau,\sigma) = -\tau
\eeq
with the last equality following from the light-cone 
gauge condition.
But in the lightcone gauge we must have $X^{\prime} 
_+(-\tau)=-\tau$ 
which is now consistent with the RHS, due to the time-parity 
reversal in the spacetime. 

We can check that $X_-$ and 
$\psi^-$ transform properly. Inspecting (\ref{gaugefix}) indicates 
that reflecting the worldsheet coordinates   
implies that $\partial _+ X^-$ is 
invariant and $ \psi^-$ flips sign. Combining 
these with the invariance of $\partial_+ X^+$ and 
$\psi^+=0$, one finds that 
a reversal of 
the worldsheet coordinates, 
while  {\em staying within the light-cone gauge}, 
requires $X^{0,1}$ and 
$\psi^{0,1}$ to also change sign. 
Assuming diffeomorphism invariance of the quantized string, 
this action 
is precisely 
$\theta$.

In ten dimensions one finds the effective action
at two derivatives 
has a $PT$ symmetry. What about $C$? $C$ is trivial, since
the $E_8$ gauge group has no non-trivial outer automorphisms.
So the $PT$ symmetry is the same as $CPT$. 

\subsubsection{Discrete Symmetries of the Matrix Model}
\label{hetmatrixsec}

The matrix model describing the $E_8 \times E_8$
heterotic string or Horava-Witten theory has 
been developed by a number of authors 
\cite{kachrusilverstein,motl,hetmatrixanom,banksmotl}. It 
is given
by 1+1 dimensional
super-Yang-Mills $O(N)$ theory
with $(0,8)$ supersymmetry. The gauge supermultiplet
$(A_{\mu},\lambda_-)$ includes eight left-moving
gauginos. In addition there are eight
bosons $X^i$ and eight right-moving fermions
$\psi^i$
each in the symmetric representation, and
32 left-moving fermions $\chi^r$
in the vector representation. These fermions must be added to
cancel anomalies \cite{hetmatrixanom}.
This theory has a spin(8) R-symmetry. Under it, 
$\lambda$ and $\psi$
transform as ${\bf 8_s}$ spinors of opposite chirality, $\chi$ is
neutral and $X^i$ transforms as an ${\bf 8_v}$ vector.

The relation of this theory to the weakly and strongly
coupled limits of the heterotic theory is described
by Banks and Motl \cite{banksmotl}.
The gauge coupling of the heterotic
matrix theory is proportional to
the volume of the dual circle that the gauge theory is
compactified on. The weakly coupled heterotic string limit
corresponds to small radius in M-theory which translates
to large radius in the Yang-Mills description. In two dimensions
the gauge coupling is a relevant parameter, so this is a
strong coupling limit. On the moduli space 
the lowest energy excitations are
diagonal matrices, which gives eight bosons, 
eight right-moving fermions and 32 left-moving 
fermions.  The $O(N)$
gauge theory is completely broken to a number of $Z_2$
subgroups (which act trivially on the $X$'s). 
The truncation of the matrix
theory to these states reproduces the free heterotic string
theory.

The Lagrangian is
given by \cite{hetmatrixanom}
\ba
L &=& \hbox{tr} \left(i \lambda D_+ \lambda + \hbox{Tr} F^2_{\mu \nu} +
D^{\mu} X^i D_{\mu} X^i
+i \psi^i D_- \psi^i + i \chi^r D \chi^r \right.
\nonumber \\ 
& &\left. -(X^i T^a X^j)^2 + i \psi \tilde{\gamma}^i \lambda
X^i \right) ~.
\ea
Here $\tilde{\gamma}^i$ are eight dimensional 
matrices and are all real 
(but not all symmetric). 
They are defined such that $\gamma^i \equiv 
\tilde{\gamma}^i \otimes \sigma_1$
satisfy the spin(8)
Clifford algebra (with positive signature).

There is one symmetry $\theta_M$ which is not part of 
the spin(8) R-symmetry. It includes a 
reversal of the 1+1 dimensional 
coordinates and is given by  
\ba
A^{a}_{\mu} (z)
&\rightarrow& -A^{a}_{\mu}(-z) ~,~  \chi_a(z) 
\rightarrow \chi_a(-z)
\nonumber \\
X^{i}_{ab}(z,\bar{z}) &\rightarrow&   X^i_{ab} (-z,-\bar{z}) 
\nonumber
\\
\lambda_{ab}(z) & \rightarrow & - \lambda_{ab}(-z) ~,~
\psi_{ab}(\bar{z}) \rightarrow  \psi_{ab}(-\bar{z}) 
\ea
where $a,b$ are $O(N)$ color indices, and 
the spin(8) 
index has been suppressed on the spinors. 
Note that there is a crucial $(-)$ sign appearing in the 
transformation of the gauginos $\lambda$. This is needed 
to make the 
Yukawa coupling invariant. (And in our notation 
$T^{-1} c \theta \eta T= c^* T^{-1} \theta T T^{-1} \eta T$.)

The transformation 
$\theta_M$ acts to send $L(z) \rightarrow L(-z)$ 
which is sufficient for it to be a 
symmetry of the theory.
To verify this, it is important to remember to 
complex conjugate all 
constants appearing in the Lagrangian. 
This includes
the generators appearing in both the
covariant derivatives and explicitly
in the Lagrangian. For all
the $O(N)$ representations
appearing here 
the generators
are purely imaginary. (So that, for
example,
$A^a T^a$ is invariant under $\theta_M$.)
 
On the moduli space all the gauginos and gauge bosons
are massive. Restricting attention to just the 
zero modes, the action becomes   
the heterotic string theory 
action in the light-cone gauge. Since the 1+1 
dimensional spinors are in the spinor representation of 
spin(8), the theory we arrive at is  
the Green-Schwarz formulation 
of the heterotic theory.
The action
of the $\theta_M$  symmetry on the diagonal zero
modes
$X^j$,
$\psi^j$ and the 32 $\chi^r$, is rather trivial, for 
it only reverses the worldsheet coordinates, without 
any additional transformation on the fields.
Transforming to 
the RNS formulation, 
this symmetry still reverses only the world-sheet 
coordinates.  
As described in the previous section, this is 
just $\theta$ in the light-cone gauge.

Thus $CPT$ of the weakly coupled 
heterotic theory is also a symmetry of
its non-perturbative formulation. This also
establishes that $CPT$ is a symmetry of the Horava-Witten
theory. Again, as before our arguments do not 
apply to backgrounds with M5 (NS5) branes. 

\subsection{M-Theory on $T^5$}
\label{m5torus} 

By Seiberg's argument this is equivalent to the $(2,0)$
little
string theory on the dual five--torus. The little string
theory is not a local quantum field theory. So we cannot
use this non-perturbative formulation to argue
that this theory has a $CPT$ symmetry. Discrete Light Cone
Quantizations of this theory exist and are given by 1+1
dimensional conformal field theories 
\cite{seibergrozaliberkooz} . We leave it to future work 
to decide whether these theories preserve $CPT$.
Given that so little is known about little string theories, 
we can not make a reliable statement here.

\section{Conclusions}

We have conducted a search for anomalies in discrete symmetries in
a variety of models.  
In both asymmetric and symmetric 
orbifold models, with and
without supersymmetry, we have not found discrete anomalies.
This perhaps might be surprising,
given that asymmetric orbifolds are inherently more
``stringy'' than symmetric orbifolds. 

We have also
seen that the {\em untwisted} sector often
has non-universal anomalies. But in string
theory the states in the {\em twisted} sector are
also charged under these symmetries.
This
is not surprising from string theory, but it is from
effective field theory. The discrete symmetries we are
studying are, after all, discrete isometries of the internal
manifold (in this case, an orbifold). It is only after
adding the contributions from the untwisted and twisted
sectors are the anomalies universal and may be canceled
by a shift of the universal axion.

In asymmetric
orbifolds a similar pattern was found, but
there is an additional subtlety. Here the charge
assignments in the twisted sector involved a
delicate
correlation  between the gauge quantum numbers of
the left--movers and the bosonic twist operators of the right--movers.
Only after properly including this correlation are
universal anomalies obtained.

We also looked for discrete anomalies in $N=1$ Type IIB
orientifold models. Here non-universal anomalies were
found, but since the world-sheet parity projection
breaks the quantum symmetry of the orbifold,
states in the twisted sector can now couple to 
$F \tilde{F}$. As a result, 
a non-universal discrete anomaly may be canceled
by assigning a shift to a twisted scalar.  In fact,
in all of the examples we studied, we found that by
including in the discrete symmetry a discrete gauge 
transformation, unbroken, non-anomalous discrete
symmetries could be identified.

We also examined several matrix model descriptions of 
M-theory compactifications and found evidence for 
CPT conservation. This would be non-perturbative 
evidence that the CPT symmetry 
of the classical theory  is preserved 
by quantum gravity. 
Our arguments are not complete, however, since 
our backgrounds did not include all types of 
five-branes. We did not consider the AdS-CFT
correspondence.  Clearly in this case, the
CPT of the boundary field theory establishes
a corresponding symmetry of boundary correlators.
Again, this is suggestive that CPT is always
a non-perturbative symmetry of string/M theory.

In undertaking this work, we were hoping that one might find
anomalies in the case of non-supersymmetric strings, but no
anomalies for supersymmetric ones.  This might have provided
evidence that string theory ``prefers" supersymmetry.  But from
this perspective, there seems to be no difference between
supersymmetry and its absence.

\noindent

{\bf Acknowledgements:}

\noindent

This work supported in part by a grant from the U.S.
Department of Energy.  Some early stages of this analysis
were conducted with K. Choi and J. Gray.  We thank Tom Banks, K. Choi, 
J. Gray, Lubos Motl, F. Quevedo, 
J. Polchinski, J. Preskill and E. Silverstein for many
helpful discussions. MG thanks the Aspen Center for Physics 
where part of this work was completed. Preprint numbers  
CALT-68-2519 and SCIPP-04/17.

\section{Appendix A: Review of Symmetric Orbifolds with Wilson Lines }
\subsection{Construction of Models} 

We begin
with a brief discussion of the construction of
symmetric orbifolds, with or without Wilson lines \cite{orbifold12}. See 
for instance, \cite{ibanez} for a more detailed 
discussion. 

We consider compactification of heterotic strings
on a six-dimensional lattice with periodic 
boundary conditions. 
At special points in the
moduli space the lattice 
may have a point group of
symmetries. 
In addition to the point group the torus has a
larger space group, which includes invariance under
translations by lattice vectors $e^i_{\alpha}$ :
\beq
X^i \rightarrow X^i + m^{\alpha}  e^i_{\alpha} ~.
\eeq

The construction of the orbifold proceeds by first
selecting an element of the point group
\beq
\theta = e^{ 2\pi i \vec{\phi} \cdot \vec{R}}
\eeq
which acts on the spacetime variables $X^i$. This
is identified
with an element
\beq
e^{2 \pi i \vec{\beta} \cdot \vec{T} }
\eeq
of the $SO(32)$ or $E_8 \times E_8$
gauge group. 

To add Wilson lines one
embeds the space group into the gauge group. This is
done by
associating a translation element $e_{\alpha}$ with an
action by the gauge group:
\beq
\lambda \rightarrow \gamma(e_{\alpha}) \lambda ~.
\eeq
We denote
\beq
\gamma(e_{\alpha})= e^{2 \pi i \vec{a} _{\alpha} \cdot \vec{T}}
\eeq
and the $\vec{a}_{\alpha}$ are referred to as Wilson lines.
The orbifold group $g$ then consists of the elements
$(\theta,0 ; \beta (\theta))$ and $(1,e_{\alpha},\gamma (e_{\alpha}))$.
Loosely speaking, the states of the theory are 
those that are invariant under this group.

There are constraints on the
allowed values of the Wilson lines. In addition to the requirement
that $N \vec{\phi}= N \vec{\beta}=
N \vec{a}_{\alpha} = 0 $ mod 1 which follows from the
requirement that $g^N=1$ ($N$ is the order of the
group) , Wilson lines also satisfy \cite{ibanez}
\beq
\gamma(\theta e_{\alpha}) = \beta ^{-1}(\theta)
\gamma ({e _{\alpha}}) \beta
(\theta)          ~.
\eeq
This follows from requiring that the multiplication law
for the gauge group elements is
homomorphic to the multiplication law for the
space group.
For abelian embeddings it reduces to
$\vec{a}_{\theta e_{\alpha}} = \vec{a}_{\alpha} + \vec{n} $.
For the $Z_6$ twists that we study this constraint is enough
to forbid any Wilson lines on the $Z_6$ twisted planes.
A Wilson line in a $Z_3$ twisted plane is allowed provided
that $3 a_i=0$ mod 1.

The level matching conditions, necessary for 
modular invariance, with a Wilson line
are
\beq
(\beta + m^{\alpha} a_{\alpha})^2 - \phi^2 = 0 ~\hbox{mod}~ 2N
~;~ N \hbox{tr} a_{\alpha} = 0 ~\hbox{mod} ~2 ~,
N \hbox{tr} \phi=0 ~\hbox{mod} ~2 ~.
\eeq
where $N$ is even. 

States in the untwisted sector are obtained by projecting
onto states that are invariant under $(\theta, 0;\beta(\theta))$.
If in addition
Wilson lines are present, untwisted states
must also be invariant
under $(1,e_{\alpha},\gamma (e_{\alpha}))$.

States in the twisted
sectors are obtained by projecting onto states invariant 
under the point group, consistent 
with the GSO projection. 
When Wilson lines are present there is one additional 
rule - one only keeps
states invariant
under
those space group elements that commute with the
twist for that sector.

Here though
there a few subtleties in obtaining the orbifold group 
transformation 
property of the twisted 
(worldsheet) 
bosonic and fermionic ground states. 
In particular, the fixed points will in general 
not be invariant, but will transform into one another. 
Fixed points with well--defined charge are obtained 
simply by diagonalization.

The other subtlety is to correctly
obtain the orbifold transformation of
the world--sheet 
fermion ground states. This is straightforward in the
bosonic formulation since, for instance,
 the Ramond right-mover (RM) ground state
corresponds to the twist operator
\beq
\tau_R = e^{-i ([n\phi_R]-1/2) \cdot H_R} ~,
\label{twistedramond}
\eeq
where $0 \leq [\xi ] <1$. Under $g$
\beq
H_R \rightarrow H_R + 2 \pi \phi
\eeq
from which the transformation of the twist operator under 
$g$ is found. 
The charge of the left-mover (LM) fermions is obtained in a similar way.

\subsection{Discrete Charges} 

In general the charges of the worldsheet 
fermions 
are more easily obtained by bosonization. 
In the
untwisted sector the RM Ramond ground state
is described by the vertex operator
\beq
e^{-i r \cdot H_R}
\label{rmgroundstate}
\eeq
where $r$ is an $SO(8)$ spinor $(\pm,\pm,\pm,\pm)/2$
and the GSO projection
requires an even number of $+$'s. The charge 
of this state under a $Z_k$ rotation
of the third plane $H^3_R \rightarrow H^3 _R + 2\pi q/k$,
for instance, is then obtained 
directly from
(\ref{rmgroundstate}). The contribution from
any oscillator
are obtained rather trivially. The total charge of 
a physical state built from oscillators and 
the R ground state is then easily obtained. 

In the twisted sectors both the fermionic ground state and
the bosonic fixed points contribute a
charge. The former is obtained from the bosonized
Ramond sector twist operator (\ref{twistedramond}).
Under a $Z_k$ rotation 
where the third torus transforms as above the 
discrete charge of this operator is 
obtained rather straightforwardly. 

The other non-trivial contribution involves the
bosonic ground states, which 
are also charged under
these discrete symmetries. This is understandable, 
since the fixed points themselves 
transform. 
For example, consider the $Z_6$ orbifold
with twist $(1,1,-2)$/6. In the $n=1$ sector there
are 3 fixed points coming from the third torus.
They are invariant under
simultaneous
$Z_6$
rotations $\eta$
of the first two tori. But under a $Z_6$ rotation $\gamma$ of
the third torus they transform as $3=2 + \gamma^3$.
In the doubly twisted sector the
27-fold degeneracy transforms as
\beq
27 \rightarrow 10(1;1) + 5(1; \gamma^3) + 8(\alpha^3; \eta^3) + 4(\alpha^3;
\gamma^3\eta^3)
\eeq
where the phase $\alpha$ $(\alpha^6=\eta^6=1)$  under the
orbifold transformation $(g)$ has also
been included.
In the triply twisted sector the 16-fold degeneracy transform
as
\beq
16 \rightarrow 6(1;1) + 5(\alpha^2; \eta^2) + 5(\alpha^4; \eta^4)  ~.
\eeq

Putting these elements together gives the total 
discrete charge of a state.

\section{Appendix B : Review of Asymmetric Orbifolds}

Asymmetric orbifolds are described by a 
set of fields valued in an internal lattice (described 
below) 
together with a specification of the orbifold 
group action, consistent with modular invariance. 
The fields in the theory are 
the 16 freely
interacting left--moving (LM) real scalars $H^a_L$, 
three LM complex scalars $X^i_L$, three right--moving 
(RM) complex scalars 
$X^i_R$ and their fermionic partners 
$\tilde{\psi}_i$. The scalars $X$ cannot be interpreted 
as describing the coordinates of an internal 
manifold as the left and right movers are 
treated differently.

To construct an asymmetric orbifold one first
begins with a Lorentzian, even, self-dual lattice $\Lambda_L \equiv
\Gamma_{(22,6)}$. These conditions are
required by modular invariance and consistency of operator
products. The lattices we consider are of the form
$\Gamma_{(22,6)}= \Gamma_{(16)} \times \Gamma_{(6,6)}$. 
Both the $SO(32)$ and $E_8\times E_8$
lattices are self-dual and even. For the non-supersymmetric 
models we consider  
the $SO(32)$ lattice and for the supersymmetric models 
we consider both. 

A simple construction \cite{ao} of a lattice $\Gamma_{(6,6)}$
that
satisfies these properties is to begin with a
Lie algebra ${\cal G}$ and to consider
momenta in the weight lattice of ${\cal G}$,
\beq
p_L,p_R \in \Lambda_W ~.
\eeq
The weight lattice is the integer
sum of all weights of ${\cal G}$.
The weight lattice can be
decomposed into cosets
\beq
\Lambda _W =
 \uplus   _a (\Lambda _R + w_a) ~,
\eeq
with one $w_a$ from each conjugacy class.
A Lorentzian lattice $\Lambda_L({\cal G})$
is formed by the elements $(p_L,p_R)$,
where the inner product
is chosen with the
appropriate signature. With
the difference of the left and right momenta restricted
to lie in the root lattice,
\beq
p_L - p_R \in \Lambda_R ~,
\eeq
it follows that the Lorentzian lattice is both integral, $l \circ
l^{\prime}
=p_L \cdot p^{\prime} _L  - p_R \cdot p^{\prime}_R
= 0 ~\hbox{mod} ~1$,
and even,
$l
\circ l
=0
~\hbox{mod}
~2$. If we further choose both $p_L$ and $p_R$ from the
same conjugacy class, then the lattice
is also self-dual \cite{phyrep}.

These models still have 16 supercharges. To obtain 
nonsupersymmetric and $N=1$ supersymmetric models 
we need to orbifold. 
This is done, as before, by dividing out by an abelian point group
symmetry of the lattice.
One set of symmetries of the lattice
consists of the Weyl groups $g$ of ${\cal G}$. From
the construction of the lattice we see
that there is an
independent Weyl symmetry for both the left and right movers.
In an asymmetric
orbifold one may choose to twist only the left or
the right by these Weyl symmetries.

The full lattice may have
additional
discrete symmetries that are not Weyl symmetries.
There are, for instance, symmetries
which interchange the conjugacy classes.

In the remainder of this 
Appendix we provide the orbifold group action, the mass 
formulae of states, and the constraints provided 
by level matching. 

The scalars $H^a_L$ have
momenta on
the $SO(32)$ lattice $\Gamma_{(16)}$.
The orbifold action on these fields may include a shift
\beq
H^a_L \sim ~H^a_L + 2 \pi \beta^a_L ~.
\eeq

The scalars $(X^i_L, X^i_R)$
are valued in the torus $R^{6,6}/\Gamma_{(6,6)}$. 
A twist or shift is given by 
\beq
X^{i}_{L(R)} \rightarrow e^{2 \pi i \phi^{i}_{L(R)}} X^{i}_{L(R)}
+ \beta^i _{L(R)} ~.
\eeq
By worldsheet supersymmetry there is an action on the right-moving
fermions:
\beq
\tilde{\psi}_i \rightarrow e^{-2 \pi i \phi^i _R} \tilde{\psi}_i ~.
\eeq

To describe the right-moving fermions $\tilde{\psi}^a$,
$a=1,..8$
and obtain the projectors
it is more convenient to bosonize the
right--moving fermions to four real scalars $H^a _R $, $a=1,..4$.
The relation between the fermions and bosons is given by
$\tilde{\psi} ^a  = e^{-i H^a _R}$ for the Neveu-Schwarz
sector, and $\tilde{\psi} ^a  = e^{\pm i {1 \over 2} H^a _R}$ for
the Ramond sector. The GSO projection
requires the momentum
$r^a$ of $H^a _R$ to be from either
the
vector or one of the spinorial weight lattices of $SO(8)$, for the
Neveu--Schwarz or Ramond sectors respectively.
The group action of the orbifold
is given by 
\beq
H_R \rightarrow H_R + 2 \pi \phi_R  ~.
\eeq
The
momentum in the twisted sectors is then given by $r+n\phi_R$.
Twisted states without oscillators
correspond to
the fermion twist operator
\beq
\tau_R \sim e^{-i (r + n \phi_R ) \cdot H_R } ~.
\eeq
States with oscillators correspond to multiplying this operator
by the appropriate factors of $\overline{\partial} ^n H_R$.

Finally, the mass formulae for the right and left movers in
the $n$--th twisted sector are respectively
\ba
E_R &=& \overline{N}_{osc} + {1 \over 2} (r + n \phi_R)^2
+ {1 \over 2} (p_R + n \beta_R)^2 +  h_R  - {1 \over 2} ~,
\label{energyr}
\\
E_L &=& N_{osc} + {1 \over 2} (p_L + n \beta _L)^2 +h_L -1 ~.
\ea
Here $N_{osc}$ and $\overline{N}_{osc}$ are the
number operators for the left and right
moving oscillators. Also,
\beq
h_{L(R)} = {1 \over 2} {k_{L(R)} \over N}(1 - {k_{L(R)} \over N})
\eeq
where $0 \leq k/N <1$, 
are the shifts in the zero point energies due to the twisted
oscillators and contains an implicit sum over all the 
twisted bosons.

Level matching provides a constraint on the
allowable left--mover shift $\beta_L$ and twists $(\phi_L,\phi_R)$,
and is necessary for maintaining one-loop modular invariance.
For $N$ even the constraints are \cite{ao}
\beq
N (\beta _L \cdot \beta_L   - \phi_L \cdot \phi_L
- \beta_R \cdot \beta_R )
=   0 ~\hbox{mod 2}
\label{lm1}
\eeq
\beq
N \sum_i  \phi^i _R  =0 ~\hbox{mod} ~2
~,  ~ N \beta^a_L  \in \Lambda_R ~, ~N \beta_L \in \Lambda _R~,
\label{lm3}
\eeq
\beq
p_L \cdot \theta_L ^{N/2} p_L
- p_R \cdot \theta _R ^{N/2} p_R =0 ~\hbox{mod} ~2 ~.
\label{lm2}
\eeq
For $N$ odd (\ref{lm3}) is still required
but (\ref{lm2}) is not,
and
(\ref{lm1}) must be satisfied mod 1.
In \cite{vafafreed} Freed and Vafa
prove that for abelian symmetric orbifolds,
these level matching conditions are necessary and sufficient
to guarantee higher loop modular invariance. It is
implied that a similar statement is also true for
abelian  asymmetric orbifolds (see footnote 2 of \cite{vafafreed}.)

Modular invariance requires the states to
satisfy certain projection rules. These may be obtained
from the twisted partition
functions, which are found
by applying a sequence of modular transformations
to the untwisted partition function.
We checked that the partition
functions obtained this way 
satisfied the following
important self-consistency condition. 
Starting from a partition function $Z_{(m,n)}$
twisted some number of times
in the two directions of the torus, perform a sequence
of modular transformations that close to
the identity.
Then one should find that $Z_{(m,n)}$ is invariant under
this action, which is indeed
confirmed by explicit computations.

\subsection{Level Matching Condition in Asymmetric $Z_6$ Models} 

For the non-supersymmetric models we 
considered the twist 
\beq 
\phi_R=({1 \over 3}, {1 \over 2}, {1 \over 2}) ~. 
\eeq 
Here we show that this twist satisfies the last 
level matching condition (\ref{lm2}). 
To see this,
write $p=p_{(2)} + p_{(4)}$ with
$p_{(2)} \in \Lambda_W (SU(3))$ and $p_{(4)} \in \Lambda_W (SO(8))$.
For then
\ba
p_L \theta^3 p_L - p_R \theta ^3 p_R &=& p^{2}_{(2),L}-p^{2}_{(2),R}
+p^{2}_{(4),L} + p^{2}_{(4),R} \nonumber \\
& = & p^2_L -p^2_R +2 p^2 _{(4),R} ~.
\ea
Now the first two terms are just $l \circ l$ which is
even (since the lattice is even).
The last term is also even, since
$p^2_{(4),R} = 0 ~\hbox{mod}~1$
for any weight of the $SO(8)$ lattice. It follows
that the right side
of the above equation is even as required.

For the supersymmetric models we considered 
\beq
\phi_R=(-1/3,1/6,1/6)  ~~; ~~
\phi_L=(0,1/6,1/6) ~.
\eeq
We now see that the last level matching condition is
satisfied. With $(p_{(i),L},p_{(i),R}) \in \Gamma^{i} _{(2,2)}(A_2)$,
\ba
p_L \theta^3 p_L - p_R \theta^3 p_R  &=& p^2_{(1),L}- p^2_{(1),R}
-\left(p^2_{(2),L} - p^2_{(2),R} + p^2_{(3),L} - p^2_{(3),R}
\right) \nonumber
\\
& = &  - p \circ p + 2 ( p^2_{(1),L}- p^2_{(1),R})  \nonumber  \\
& =&    - p \circ p + 2 r \cdot (p_{(1),L}+p_{(1),R})
\ea
where $r$ is the root vector $p_{(1),L}-p_{(1),R}$.
Both terms in the last line above are manifestly even.

\section{Appendix C: Discrete Charges of Bosonic and Fermionic 
Twist Operators in Asymmetric Orbifolds} 

\subsection{Discrete Charge of 
World--Sheet Fermions}
\label{fermicharge}

It is easiest to begin with the 
charges of the worldsheet fermions.
We focus on a $Z_M$ symmetry that acts on the right--moving
bosonized NS fermions as
\beq
H^a _R \rightarrow H^a _R + {k _a \over M}  ~.
\label{Htrans}
\eeq
In the
Ramond sector the fermions are
half-integer moded, in which case
the $Z_M$ symmetry is realized as a
$Z_{2M}$ symmetry.
This is obvious from the bosonized expression for these states,
\beq
\tilde{\psi}_{R} \sim e^{- i r \cdot H_R} ~,
\eeq
since $r$ is half--integer valued.

A generic single particle state from a
twisted sector
involves
oscillators and/or lattice
momenta acting on the ground state of that sector.
The charge of the twisted 
ground state is obtained by inspecting  
the 
fermion twist operator of the $n$--th twisted
sector, which is
\beq
\tau_R \sim e^{-i (r +n \phi_R) \cdot H_R }    ~.
\eeq
Here $r$ is the momentum of the R or NS ground state in the 
$n$--th twisted
sector. Given
this explicit construction
it is straightforward to determine the $Z_M$ charge of the
worldsheet fermions in the twisted sectors
using the transformation (\ref{Htrans}). In particular,
the charge is
\beq
e^{-2 \pi i (r+ n \phi_R) \cdot {k \over M} } ~,
\label{tfrcharge}
\eeq
which
is not a multiple of $1/M$. Actually, in the twisted sectors
the
$Z_M$ symmetry is instead $Z_{2NM}$!

It is not unreasonable that the worldsheet fermions
have charges that are $Z_{2NM}$ :  this
is due to the quantum $Z_N$ symmetry.
For example, the operator product of $N$ singly twisted fermion
operators does not contain the identity element, but rather contains
an untwisted fermion. This is because the product
conserves
the quantum $Z_N$ charge, and so describes in general 
an excited state $r^{\prime}$
of
the untwisted Hilbert space. Schematically,
\beq
(\tau_F )^N
 \sim  e^{-i r^{\prime} \cdot H_R} ~.
\label{corrtauf}
\eeq
Since an untwisted worldsheet fermion has a charge
$q/M$,
the fermion twist operator must have a charge
that is a multiple of $1/(2NM)$. 
By computing a sufficient number of correlation functions 
the $Z_M$ charge is uniquely determined up to an 
additive shift proportional to the quantum $Z_N$ charge.

Given these results, one then also expects the
bosonic twist operators to be
charged.

\subsection{Bosonic Twist Operator: 
Discrete Charges from Branch Cuts}
\label{bosonictwist1}

In order to compute the anomalies it is necessary to compute the
transformation properties of the different states.  In twisted
sectors, one must be a bit careful, since the ground states
transform.  For the fermionic part of the state, it is 
useful to bosonize the
right moving fermions, for then 
the twist acts as a shift and it is easy
to read off (as in the symmetric orbifold case) the charge of the
twist operator.  For the bosonic coordinates, we will resort to a
somewhat more indirect argument.
The basic idea is simply to note that the bosonic twist and
antitwist operators contain, in their operator product expansion,
the unit operator.  The coefficient function is determined in
terms of the dimension of these operators and is in general
not single-valued.
In correlation functions, the effect of the twist operator is to
introduce a branch cut, corresponding to the $Z_3$ charge of the
operator.  From the OPE, then, one can read off the transformation
of $\tau_R$ from the value of the branch cut.

Thus if the orbifold acts as  
$X^i_R \rightarrow e^{2 \pi i /N} X^i_R$, a bosonic
twist operator from the singly twisted sector,
$\tau _B (0)$, located at the origin
introduces a branch cut :
\beq
\overline{\partial} X^i_R (\bar{z}) \tau_B (0) \sim \bar{z} ^{-1/N}
\tau ^{\prime} _B (0) ~.
\eeq
But note that the branch cut is the same as the $Z_M$ 
(here, $Z_3$) charge 
of $X_R^i $. This is because in this example, 
the orbifold group 
contains in its product the discrete symmetry 
of this torus.

Before considering the twisted bosons, 
we note that the RM $Z_3$ charge of a world--sheet fermion
can be inferred from the branch cut in this 
way. 
For illustrative purposes we consider 
an untwisted NS fermion and a twisted R fermion, 
although the charge for  
untwisted R fermions or twisted NS fermions 
can also be obtained this way. 

$\bullet$ Untwisted NS fermion. 
The leading term in the OPE for the untwisted
NS fermion and the NS fermion
twist operator
(for which $r_{NS}=(1,0,0,0)$ and $r_{NS} \cdot \phi=0$) is :
\beq
: e^{ \pm i H_I (z)} :: e^{i (r_{NS}+ \phi) \cdot  H(0) } :
~\sim  z^{\pm \phi _I } :e^{i (r^I_{NS} + \phi^I \pm 1)H_I(0)}e^{i 
{(r_{NS}+\phi) {\widehat{\cdot}} H(0)}}:
~,
\eeq
with $\widehat{\cdot}$ denoting the inner product 
omitting $H^I$. 
Note that the power of $z$ follows from dimensional analysis. In
general it
is the difference in the dimensions of the operators 
appearing on the right side with those on the left 
side. 
The branch cut is
\beq
e^{\pm 2 \pi i \phi_I}
\eeq
which is the same as the $Z_M$ charge of the untwisted
NS fermion.

$\bullet$ Twisted Ramond fermions. 
The leading term in the OPE between the 
Ramond and NS twisted fermionic operators is 
\beq
:e^{i (-{1 \over 2} + \phi)
\cdot H(z) } e^{i (r_{NS}+\phi) \cdot  H(0)}
: ~\sim z^{(\phi -1/2) \cdot \phi -1/2} 
:e^{i (r_{NS}+\phi +\phi-{1 \over 2}) \cdot H(0) } : ~.
\eeq
(The last factor of --1/2 in the power of $z$ 
is from the untwisted field $H^0$). 
Using factorization the branch cut gives the correct charge,
\beq
e^{2 \pi i \phi_I ( \phi_I -1/2) } ~.
\eeq

Now consider the twisted bosons and focus 
on the $Z_3$ symmetry of the lattice $\Gamma^{(2)}(A_2)$ 
of the third torus. It is readily confirmed 
that  
for untwisted bosons this procedure gives the correct answer.  
As stated before, this is because of factorization 
and because the orbifold group 
contains in its product the discrete symmetry we are 
studying . 

To obtain the charge of the twist operator itself, 
we note that the  
OPE of this operator with its inverse contains
the identity:
\beq
V_B(z,\bar{z}) V^{-1}_B(0)  \sim z^{2 \Delta_L}
\bar{z} ^{2 \Delta_R} \times {\bf 1} ~.
\eeq
The scaling on the right--side is determined by conformal
invariance to be the difference of the
conformal weights of the operators
appearing on both sides of the equation.
The factor of 2 appears, since the twist operator 
appears twice on the left side. 
Here we note that 
the twist
operator is {\em not } the full vertex operator, 
but only that part of the twisted vertex operator constructed 
from the fields valued in the sublattice $\Gamma_{(2,2)}(A_2)$. 
In particular, this 
includes a contribution from the LM and RM momenta 
of the ground state, so that 
\beq
V_B \sim e^{i p_L \cdot X_L(z)} e^{ -i p_R
\cdot X_R(\bar{z}) } \tau_B(z,\bar{z})
\eeq
with $(p_L,p_R) \in \Gamma_{(2,2)}(A_2)$.
(In our models $\beta_L=\beta_R=0$ when restricted 
to $\Gamma_{(2,2)}(A_2)$. )
Also the energy shift from fields valued in this 
sublattice is  
\beq 
\Delta_L = {1 \over 2} p^2_L + h_L
\eeq
and with a similar expression for $\Delta_R$. 
Again, $h_L$ is not the full shift in energy 
due to all the twisted bosons in $\Gamma_{(6,6)}$, but only that 
part from the twisted bosons in $\Gamma_{(2,2)}(A_2)$. 
The net branch cut is then:
\beq
e^{2 \pi i (2 h_L -2 h_R + p^2_L- p^2_R)}  ~.
\eeq
Note that this branch cut vanishes for a symmetric twist :
the contribution from the zero point energy cancels
between left and right, and $p_L=p_R=0$ since both
$X_L$ and $X_R$ are
twisted.
As advertised earlier, in a symmetric orbifold the
bosonic twist operator is neutral under the discrete
symmetry. 

Now focus on our example: the RM $Z_3$ twist
in the $\Gamma_{(2,2)}(A_2)$ lattice. Then $p_R=0$ for
the massless states, $h_L=0$ and
$h_R=1/9$. Then since $p_L=0$ or $p_L \in Y_{1,2}$ (for
which $p^2_L=2/3$), the branch cuts are
\ba
 e^{2 \pi i (-2/9)} &~~\hbox{if} &~~ p_L=0 \nonumber \\
e^{2 \pi i (4/3 - 2/9)}
= e^{2 \pi i (1/9) }  &~~\hbox{if} &~~p^2_L=2/3 ~.
\ea
From this we infer 
that the $Z_3$ charges of the bosonic ground states are :
\ba
\alpha^{-2/3} &~~,& ~p_L \in Y_0 \nonumber \\
\alpha^{1/3}  &~~,&
~p_L \in Y_{1,2} ~.
\label{z3bccharges}
\ea
Notice that the charge depends on the choice of LM 
momentum $p_L$. This correlation is rather 
surprising, 
since the LM are not charged under the discrete 
$Z_3$! The naive expectation that all the right-moving ground 
state have the same charge is not correct. 
We note that 
these charge assignments (\ref{z3bccharges}) are 
crucial to obtain universal anomalies. 

Independent confirmation of these charges 
is provided in 
the next subsection, where they are 
obtained 
from a completely different 
method. Fortunately, the two methods agree.

\subsection{Bosonic Twist Operator: 
Discrete Charges from an Explicit Construction}
\label{bosonictwist2}

Recall that in the case of the
twisted fermions the charge 
assignments were easily found 
by bosonizing the fermions. 
To confirm the charges in (\ref{z3bccharges}) it would be
nice to do something similar and 
obtain an explicit form for the bosonic twist
operator. 
That is, to
express the right-moving scalars
as the exponential of another set of scalars.
Then the linear action of the
twist on the original variables
would act as a non-linear shift on the new
set of scalars.
For the $\Gamma_{(2,2)}(A_2)$ lattice with the $Z_3$ twist, 
it turns out that
an
explicit rebosonization can be found and is given by \cite{tyekz}
\beq
\overline{\partial} X^{(3)}_R = {1 \over \sqrt{3}}
(e ^{i e_1 \cdot B} + e^{i e_2 \cdot B} +
e^{i e_3 \cdot B} ) ~.
\label{reboson}
\eeq
The expression for
$ \overline{\partial}~ \overline{X}^{(3)}_R$ is given by the
hermitian conjugate. Here $e_{i=1,2,3}$ are three $SU(3)$ roots
such that $e_i \cdot e_j = -1$ for $i \neq j$. $B_I$ is a
two--component RM freely interacting real scalar. 

One may confirm that this rebosonization is 
consistent with 
the conformal weights, OPE's , 
and world-sheet statistics
expected for either of these bosons \cite{tyekz}.  

Under the
orbifold action the complex scalar $\phi$ transforms as
\beq
B \rightarrow B + 2 \pi {e _4 \over 3}
\label{phitrans}
\eeq
where $e_4$ is the $SU(3)$ root for which
$e_4 \cdot
e_1 = e_4 \cdot e_2 =1$ and $e_4 \cdot e_4 =2$. With
these definitions and transformation law
for $\phi$ one indeed finds that $\overline{\partial} X^{(3)} _R
\rightarrow \gamma  \overline{\partial} X^{(3)}_R$ and
$\overline{\partial} \overline{X}^{(3)} _R \rightarrow \gamma^2
\overline{\partial} \overline{X}^{(3)}_R $.

Using this equivalence it
is now straightforward to explicitly
determine the bosonic twist operators and
their $Z_3$ charges.
They are 
 \beq
\tau_B \sim e^{-i ( p_{\phi} + n { e_4 \over 3} ) \cdot B} ~,
\label{bosetwist}
\eeq
with $p_{\phi} \equiv p_{(2),R}$.

To find the $Z_3$ charge of the bosonic ground
state we need to determine the momentum of
the ground state
appearing in (\ref{bosetwist}).
In the untwisted sector the $B$ momentum $p_{B}$
is in the $SU(3)$ weight lattice and for the twisted 
sectors in the
shifted weight lattice.
Since this scalar is not twisted it does not
modify the zero point energy in the twisted sectors.
The scalars in the other sublattice $\Gamma_{(2,2)}(D_4)$ 
are still twisted 
though. 
The
expression (\ref{energyr}) for the RM Ramond energy in
the singly twisted sector is then
\beq
E_R = \overline{N}_{osc} +{1 \over 2} (r + \phi_R)^2
+ {1 \over 2} (p_{B} + {e_4 \over 3})^2 -{1 \over 4} ~.
\eeq
This has three ground state solutions,
one from each conjugacy class :
$p_{B}=0$,
$p_{B}=w_1$ and $p_{B}=w_2$, where $w_1$ and
$w_2$ are the two weights from the fundamental and anti--fundamental
conjugacy classes such that
$w_1 + w_2=-e_4$. The three twist operators corresponding
to these three solutions are
\beq
e^{-i e_4 \cdot B /3} ~~: ~p_{B}=0 ~~; ~ e^{i e_1 \cdot B /3}
~~, ~p_{B}=w_1 ~~; ~ e^{i e_2 \cdot B/3} ~~, ~ p_{B}=w_2 ~.
\eeq
The conformal weight of 
these operators is $1/9$ which is correct, 
since in the other twisted formulation 
the weight is  
equal to the shift in the
zero point energy which in that case
is also $1/9$.

The $Z_3$ charges of the twist operators
are easily obtained
using the explicit transformation (\ref{phitrans}).
They are : $\alpha ^{-2/3}$ , $\alpha^{1/3}$,
$\alpha ^{1/3}$, respectively. Note that the charge
assignments are not identical but
depend on the choice of
RM bosonic
ground state. This surprising result was 
also found in the method of the 
previous subsection. 
More importantly, the charges
inferred from either method 
agree.

Since there are three RM ground states it appears that 
the bosonic degeneracy in the single twist sector is 
3. Naively this disagrees with the construction of this 
model in the previous (twisted) formulation, since there the 
degeneracy was one. There is no contradiction though, 
since the three RM ground states are not all paired
with each LM ground state. In fact, only one RM state is selected
according to the
correlation $p_{(2),L}-p_{B} \in R$ and the choice of $p_{(2),L}$.
That is, $p_{(2),L} \in Y_0 \leftrightarrow p_B=0$ and 
$p_{(2),L} \in Y_{1,2} \leftrightarrow p_B=w_{1,2}$.
Thus the bosonic degeneracy remains equal to one, consistent
with the degeneracy found in the twisted formulation.

To verify that the states are the same 
in either formulation and that the degeneracies 
match,  we need
to look at the projectors. One finds 
that each
LM state is always paired up with only one of the three
RM bosonic ground states. And as they should be, one 
finds that the projectors
are the same in either formulation.

Using the explicit expression for the twist operator, 
the total $Z_3$ charge, bosonic plus fermionic, in the singly--twisted 
sector is 
\beq
\alpha^{-1/2} ~:  ~~ p_L \in Y_0  ~~~; ~~~ \alpha^{1/2} ~:
~~ p_L \in Y_{1,2} ~.
\eeq
Thus the discrete charge of the twisted state depends
on the left--moving quantum numbers and their 
values agree with
those inferred from the branch cut.
By factorization and CPT the charges in the other twist
sectors are uniquely determined. They may be found in
Table \ref{discretecharges}.

\TABLE[h]{
\begin{tabular}{|c|c|} \hline
 sector  & ~{\rm Ramond state and $Z_3$ charge}  \\ \hline
& \\
untwisted   & $ ~(r_1 :  \alpha^{-{1 \over 2}} )
\oplus (r_2 :  \alpha^{1 \over 2} ) $ \\
& \\ \hline
$n=1$  & ~{\rm no states }  \\ \hline
& \\
$n=2$ & $~(p_{(2),L} \in Y_0 :  \alpha^{1 \over 2})
\oplus (p_{(2),L} \in Y_{1,2} :  \alpha^{-{1 \over 2}})$ \\
& \\ \hline
& \\
 $ n=3 $ & $~\alpha^{1 \over 2} $  \\
& \\ \hline
& \\
$n=4$ & $~(p_{(2),L} \in Y_0 :  \alpha^{-{1 \over 2}})
\oplus (p_{(2),L} \in Y_{1,2} :  \alpha^{1 \over 2}) $  \\
& \\ \hline
& \\
$n=5$ & $~(p_{(2),L} \in Y_0 :  \alpha^{1 \over 2})
\oplus (p_{(2),L} \in Y_{1,2} :  \alpha^{-{1 \over 2}})$  \\
& \\ \hline
\end{tabular}
\caption{$Z_3$ charges for positive helicity
massless fermions.}
\label{discretecharges}}

\section{Appendix D: Open String 
States in Type IIB Orientifolds}

A general discussion for constructing 
the states in a ${\cal N}=1$, D=4 Type IIB orientifold 
can be found in \cite{ibanezdbrane1,ibanezdbrane2}. 
We briefly summarize their main results that are 
relevant here.

The orientifolds studied here are obtained 
by first compactifying Type IIB on $T^2 \times T^2 \times T^2$ 
where each torus 
has a $Z_6$ isometry. The orientifold group is 
generated by the group
elements $G_1=g$ and $G_2 = g \times \Omega$ with 
$g^N=1$. Here $\Omega$ is 
the world-sheet parity transformation, 
and $g$ generates
a discrete isometry
of the internal manifold and also a discrete $U(16)$
transformation of the open string endpoints. 
The spacetime
embedding of $g$ is chosen
to be an element of $SU(3)$ so that 
at low energies these models describe ${\cal N}=1$ supersymmetric 
theories in four dimensions.

Since $G_2=G_1 \times \Omega$, states in the
closed string sector are obtained by first projecting onto $g$-invariant
states and
then onto $\Omega$-invariant states.
In the massless untwisted bosonic sector 
the NS-NS antisymmetric two-form,
R-R zero form and R-R four form states are discarded, 
but the 
dilaton, graviton and R-R two form states are kept. 
The twisted sector initially contains 
states obtained by the standard orbifold construction. 
But the world sheet parity
transformation exchanges states in the $n$--th twisted
sector with those in the $(N-n)$--th twisted sector. Only
the symmetric combination survives the world-sheet 
parity projection. As a result these models 
do not have a quantum $Z_N$ symmetry. 

As is well-known, in these models the charged matter 
is from the open string sector which are all untwisted.  
All the models here have D9 branes. They will 
also have D5 branes when the orbifold group contains
an element of order 2.
Open strings consist of
sectors that
are distinguished by whether their endpoints have
Dirichlet or Neumann boundary conditions.
So there will in general be 99,55 and 95 open strings.

$\bullet $ Massless fermions in the 
99 sector are of the form
\beq
\lambda^i_{ab} |ab \rangle \times |r^i \rangle ~.
\label{99matter}
\eeq
Here $\lambda$ is a Chan--Paton matrix for 
$U(16)$ and $r=(\pm\pm\pm\pm)$ is an $SO(8)$ spinor. The
GSO projection requires an even number of $-$ signs.
The orbifold action
on the spinor $r$ is given by an element of the $SU(4)$ subgroup.
The embedding of the spacetime twist into the gauge group is,
for abelian orbifolds, described by a shift vector $v$ which
is a Cartan element of $U(16)$. Conditions on the Chan-Paton
matrices and the shift vectors to ensure tadpole cancellation
are derived in \cite{ibanezdbrane1}. The orientifold 
projection keeps all states that are
separately invariant under
the orbifold action $g$ and the world-sheet parity 
projection.

$\bullet$ D5 branes are present 
in the $Z_6$ model discussed in the main text, but not in the
$Z_3$ models. The $5$ branes are 
assumed to be wrapped around the third torus, and 
are referred to as $5_3$ branes. 
The shift vector for the D5 branes could
in principle be different from the D9 branes, 
although here they are chosen to be identical. 
The matter content of the
55 strings are
of the form (\ref{99matter}).

$\bullet$ The new ingredient for the 59 strings is that the bosonic
modes perpendicular to the $5$--brane direction have
half-integer moding. This is due to the mixed 
Dirichlet and Neumann boundary conditions. 
In the
NS sector supersymmetry implies that
the superconformal partners ${\psi}^i$, with
$i$ not one of the 
$5_3$ brane directions,
have integer moding. These worldsheet fermions
have zero modes. Consequently, spacetime scalars are of the form
\beq
\lambda_{ab} |ab \rangle |s_1 s_2 \rangle ~.
\label{95matterb}
\eeq
The GSO projection is $s_1=s_2$, i.e. an even number of $-$ signs.
In the R sector the $\psi^i$'s, with $i$ orthogonal to
the $5_3$ brane, have half-integer moding and do
not contribute any zero modes. In this sector the only
zero modes are from worldsheet fermions with indices parallel
to the $5_3$ brane. This leads to space-time
fermions of the form
\beq
\lambda_{ab} |ab \rangle |s_0 s_3 \rangle ~.
\label{95matterf}
\eeq
The GSO projection requires an odd number of $-$ signs.
The orientifold projection
keeps the symmetric combination of 95 and 59 strings.


\end{document}